\documentclass[aps,pra,twocolumn,showpacs,groupedaddress,floatfix]{revtex4}

\usepackage{graphicx}
\usepackage{amsmath,amssymb}
\usepackage{times,txfonts}

\begin{document}

\title{Quasiparticles in Neon using the Faddeev Random Phase Approximation}

\author{C. Barbieri}
\affiliation{Gesellschaft f\"ur Schwerionenforschung, Planckstr. 1, D-64291, 
Darmstadt, Germany}
\author{D. Van Neck}
\affiliation{Laboratory of Theoretical Physics, Ghent University, 
Proeftuinstraat 86, B-9000 Gent, Belgium}
\author{W.H. Dickhoff}
\affiliation{Department of Physics, Washington University, St. Louis, MO 63130, USA} 

\date{\today}

\begin{abstract}

The spectral function of the closed-shell Neon atom is computed by 
expanding the electron self-energy through a set of Faddeev equations. 
This method describes the coupling of single-particle degrees of 
freedom with correlated two-electron, two-hole, and electron-hole pairs. 
The excitation spectra are obtained using the Random Phase 
Approximation, rather than the Tamm-Dancoff framework employed in the 
third-order algebraic diagrammatic contruction [ADC(3)] method. 
The difference between these two approaches is studied, as well as the 
interplay between ladder and ring diagrams in the self-energy.  
Satisfactory results are obtained for the ionization energies as well as the 
energy of the ground state with the Faddeev-RPA scheme that is also 
appropriate for the high-density electron gas.

\end{abstract}

\pacs{31.10.+z,31.15.Ar}
\maketitle

%%%%%%%%%%%%%%%%%%%%%%%%%%%%%%%%%%%%%%%%%%%%%%%%%%%%%%%%%%%%%%%%%%%%%%%%%%%%%%%%%%%%%

\section{Introduction}
\label{intro}

{\em Ab initio} treatments of electronic systems become unworkable for 
sufficiently complex systems.
On the other hand, the Kohn-Sham formulation~\cite{Koh.65} of density 
functional theory (DFT)~\cite{Hoh.64} incorporates many-body correlations 
(beyond Hartree-Fock), while only single-particle (sp) equations 
must be solved.
 Due to this simplicity DFT is the only feasible approach in some modern 
applications of electronic structure theory. There is therefore a continuing 
interest both in developing new and more accurate functionals and in studying 
conceptual improvements and extensions to the DFT framework.  In particular it 
is found that DFT can handle short-range interelectronic correlations quite 
well, while there is room for improvements in the description of long-range 
(van der Waals) forces and dissociation processes.%~\cite{aa}.

Microscopic theories offer some guidance in the development of extensions to 
DFT. Orbital dependent functionals can be constructed using many-body 
perturbation theory (MBPT)~\cite{Goe.94,Goe.05}. More recently, the 
development of general {\em ab initio} DFT~\cite{Bar.05,Mor.05} addressed 
the lack of a systematic improvement in DFT methods.  In this approach one 
considers an expansion of the exact ground-state wave function (e.g., MBPT 
or coupled cluster) from a chosen reference determinant. Requiring that the 
correction to the density vanishes at a certain level of perturbation theory 
allows one to construct the corresponding approximation to the Kohn-Sham 
potential.

 A different route has been proposed in Ref.~\cite{Van.06} by developing a 
quasi-particle (QP)-DFT formalism. In the QP-DFT approach the full spectral 
function is decomposed in the contribution of the QP excitations, 
and a remainder or background part. The latter part is complicated, 
but does not need to be known accurately: it is sufficient to have a 
functional model for the energy-averaged background part to set up a 
single-electron selfconsistency problem that generates the QP excitations.   
Such an approach is appealing since it contains the well-developed 
standard Kohn-Sham formulation of DFT as a special case, while  at 
the same time emphasis is put on the correct description of QPs, in the 
Landau-Migdal sense~\cite{Mig.67}. Hence, it can provide an improved 
description of the dynamics at the Fermi surface.  Given the close relation 
between QP-DFT and the Green's function (GF) formulation of many-body 
theory~\cite{FetWal,DicVan},  it is natural to employ {\em ab initio} 
calculations in the latter formalism to investigate the structure 
of possible QP-DFT functionals. 
In this respect it is imperative to identify which classes of 
diagrams are responsible for the correct description of the QP physics. 

Some previous  calculations, based on GF theory, have focused on a 
self-consistent treatment of the self-energy at the second 
order~\cite{Van.01,Pei.02,Dah.05} 
for simple atoms and molecules. For the atomic binding energies it was found 
that the bulk of correlations, beyond Hartree-Fock, are accounted for while 
significant disagreement with experiment persists for QP properties like 
ionization energies and electron affinities.
The formalism beyond the second-order approximation was taken up in 
Ref.~\cite{Ver.06,Shi.06,Dah.06,Dah.04,Sta.06} by employing a self-energy of 
the $GW$ type~\cite{Hed.65}. In this approach, the random phase 
approximation (RPA) in the particle-hole ({\em ph}) channel is adopted to 
allow for possible collective effects on the atomic excited states. The 
latter are coupled to the sp states by means of diagrams like 
the last two in Fig.~\ref{fig:SigR}(c).
Two variants of the $G_0W_0$ formalism were employed in Ref.~\cite{Ver.06} 
(where the subscript ``$0$'' indicates that non-dressed propagators are used). 
In the first only the direct terms of the interelectron Coulomb potential are 
taken into account. In the second version, also the exchange terms are 
included when diagonalizing the {\em ph} space [generalized RPA (GRPA)] and 
in constructing the self-energy [generalized $GW$ ($GGW$)]. Although the 
exchange terms are known to be crucial in order to reproduce the 
experimentally observed Rydberg sequence in the excitation spectrum of 
neutral atoms, they were found to worsen the agreement between the 
theoretical and experimental ionization energies~\cite{Ver.06}. 

In the $GW$ approach the sp states are directly coupled with the 
two-particle--one-hole ({\em 2p1h}) and the two-hole--one-particle 
({\em 2h1p}) spaces. However, only partial diagonalizations (namely, in 
the {\em ph} subspaces) are performed. This procedure unavoidably neglects 
Pauli correlations with the third particle (or hole) outside the subspace. 
In the case of the $GGW$ approach, this leads to a double counting of the 
second order self-energy which must be corrected for 
explicitly~\cite{Fuk.64,Bar.06b}. We note that simply subtracting the double 
counted diagram is not completely satisfactory here, since it introduces poles 
with negative residues in the self-energy. More important, the interaction 
between electrons in the two-particle ({\em pp}) and two-hole ({\em hh}) 
subspaces are neglected altogether in ($G$)$GW$.
Clearly, it is necessary to identify which contributions, beyond $GGW$, are 
needed to correctly reproduce the QP spectrum. 
 
In this respect, it is known that highly accurate descriptions of the 
QP properties in finite systems can be obtained with the algebraic 
diagrammatic construction (ADC) method of Schirmer and 
co-workers~\cite{Sch.83}. 
The most widely used third-order version [ADC(3)] is equivalent to the 
so-called extended $2p1h$ Tamm-Dancoff (TDA) method~\cite{Wal.81} 
and allows to predict ionization 
energies with an accuracy of 10-20 mH in atoms and small molecules. 
Upon inspection of its diagrammatic content, the ADC(3) self-energy is seen 
to contain all diagrams where TDA excitations are exchanged between the three 
propagator lines of the intermediate $2p1h$ or $2h1p$ propagation. The TDA 
excitations are constructed through a diagonalization in either $2p1h$ or 
$2h1p$ space, and neglect ground-state correlations. 
However, it is clear that use of TDA leads to difficulties for 
extended systems. In the 
high-density electron gas {\em e.g.}, the correct plasmon spectrum requires 
the RPA in the $ph$ channel, rather than TDA. 

In order to bridge the gap between the QP description in finite and extended 
systems, it seems therefore necessary to develop a formalism 
where the intermediate excitations in the $2p1h/2h1p$ propagator are 
described at the RPA level. 
This can be achieved by a formalism based on employing a set of Faddeev 
equations, as proposed in Ref.~\cite{Bar.01} and subsequently applied to 
nuclear structure problems~\cite{Bar.02,Dic.04,Bar.06}. In this approach 
the GRPA equations are solved separately in the {\em ph} and {\em pp/hh} 
subspaces. The resulting polarization and two-particle propagators are then 
coupled through an all-order summation that accounts completely for Pauli 
exchanges in the {\em 2p1h}/{\em 2h1p} spaces. 
This Faddeev-RPA (F-RPA) formalism is required if one wants to 
couple propagators at the RPA level or beyond. 
Apart from correctly incorporating  Pauli exchange, F-RPA takes 
the explicit inclusion of ground-state correlations into account, and 
can therefore be expected to apply to both finite and extended systems.  
The ADC(3) formalism is 
recovered as an approximation by neglecting  ground-state correlations 
in the intermediate excitations ({\em i.e.} replacing RPA with TDA phonons).  

In this work we consider the Neon atom and apply the F-RPA method 
to a nonrelativistic electronic problem for the first time.
The relevant features of the F-RPA formalism  
(also extensively treated in Ref.~\cite{Bar.01}), are introduced 
in Sect.~\ref{theory}. 
The application to the Neon atom is discussed in 
Sec.~\ref{results}, where we also investigate the separate effects of 
the ladder and ring series on the self-energy, as well as the differences 
between including TDA and RPA phonons. 
Our findings are summarized in Sec.~\ref{conclusions}. 
Some more technical aspects are relegated to the appendix, where 
the interested reader can find  the derivation of the Faddeev expansion 
for the {\em 2p1h}/{\em 2h1p} propagator, adapted from Ref.~\cite{Bar.01}. 
In particular, the approach used to avoid the 
multiple-frequency dependence of the Green's functions is discussed 
in App.~\ref{app_time}, along with its basic assumptions. The explicit 
expressions of the Faddeev kernels are given in App.~\ref{app_kernel}. 
Together with Ref.~\cite{Bar.01}, the appendix provides sufficient information 
for an interested reader to apply the formalism.  

\section{Formalism}
\label{theory}

The theoretical framework of the present study is that of propagator theory, 
where the object of interest is the sp propagator, instead 
of the many-body wave function.
In this paper we will employ the convention of summing over repeated indices, 
unless specified otherwise.
Given a complete orthonormal basis set of sp states, labeled by 
$\alpha$,$\beta$,..., the sp propagator can be written in its Lehmann 
representation as~\cite{FetWal,DicVan}
\begin{equation}
 g_{\alpha \beta}(\omega) ~=~ 
 \sum_n  \frac{ \left( {\cal X}^{n}_{\alpha} \right)^* \;{\cal X}^{n}_{\beta} }
                       {\omega - \varepsilon^{+}_n + i \eta }  ~+~
 \sum_k \frac{ {\cal Y}^{k}_{\alpha} \; \left( {\cal Y}^{k}_{\beta} \right)^*  
}
                       {\omega - \varepsilon^{-}_k - i \eta } \; ,
\label{eq:g1}
\end{equation}
where ${\cal X}^{n}_{\alpha} = {\mbox{$\langle {\Psi^{N+1}_n} 
\vert $}} c^{\dag}_\alpha {\mbox{$\vert {\Psi^N_0} \rangle$}}$~
(${\cal Y}^{k}_{\alpha} = {\mbox{$\langle {\Psi^{N-1}_k} \vert $}} 
c_\alpha {\mbox{$\vert {\Psi^N_0} \rangle$}}$) are the spectroscopic 
amplitudes, $c_\alpha$~($c^\dag_\beta$) are the second quantization 
destruction (creation) operators and $\varepsilon^{+}_n = E^{N+1}_n -
 E^N_0$~($\varepsilon^{-}_k = E^N_0 - E^{N-1}_k$).
In these definitions, $\vert\Psi^{N+1}_n\rangle$, $\vert\Psi^{N-1}_k\rangle$ 
are the eigenstates, and $E^{N+1}_n$, $E^{N-1}_k$ the eigenenergies of the 
($N\pm1$)-electron system. Therefore, the poles of the propagator reflect the 
electron affinities and ionization energies.

%\begin{eqnarray}
% g_{\alpha \beta}(\omega) &=&
% \sum_n  \frac{ {\mbox{$\langle {\Psi^N_0}     \vert $}} c_\alpha      {\mbox{$\vert {\Psi^{N+1}_n} \rangle$}}
%                {\mbox{$\langle {\Psi^{N+1}_n} \vert $}} c^\dag_\beta  {\mbox{$\vert {\Psi^N_0}     \rangle$}} }
%                       {\omega - (E^{N+1}_n - E^N_0) + i \eta }  
%\nonumber \\
% & & ~+~
% \sum_k \frac{  {\mbox{$\langle {\Psi^N_0}     \vert $}} c^\dag_\beta  {\mbox{$\vert {\Psi^{N-1}_k} \rangle$}}
%                {\mbox{$\langle {\Psi^{N-1}_k} \vert $}} c_\alpha      {\mbox{$\vert {\Psi^N_0}     \rangle$}} }
%                       {\omega + (E^{N-1}_k - E^N_0) - i \eta } \; ,
%\label{eq:g1xx}
%\end{eqnarray}
%where $c_\alpha$~($c^\dag_\beta$) are the second quantization destruction (creation) operators. In Eq.~(\ref{eq:g1}),
%$\vert\Psi^{N+1}_n\rangle$, $\vert\Psi^{N-1}_k\rangle$ are the eigenstates, and $E^{N+1}_n$, $E^{N-1}_k$ the eigenenergies of the ($N\pm1$)-electron %system. Therefore, the poles of the propagator reflect the electron affinities and ionization energies.

\begin{figure}
\includegraphics[width=\columnwidth,clip=true]{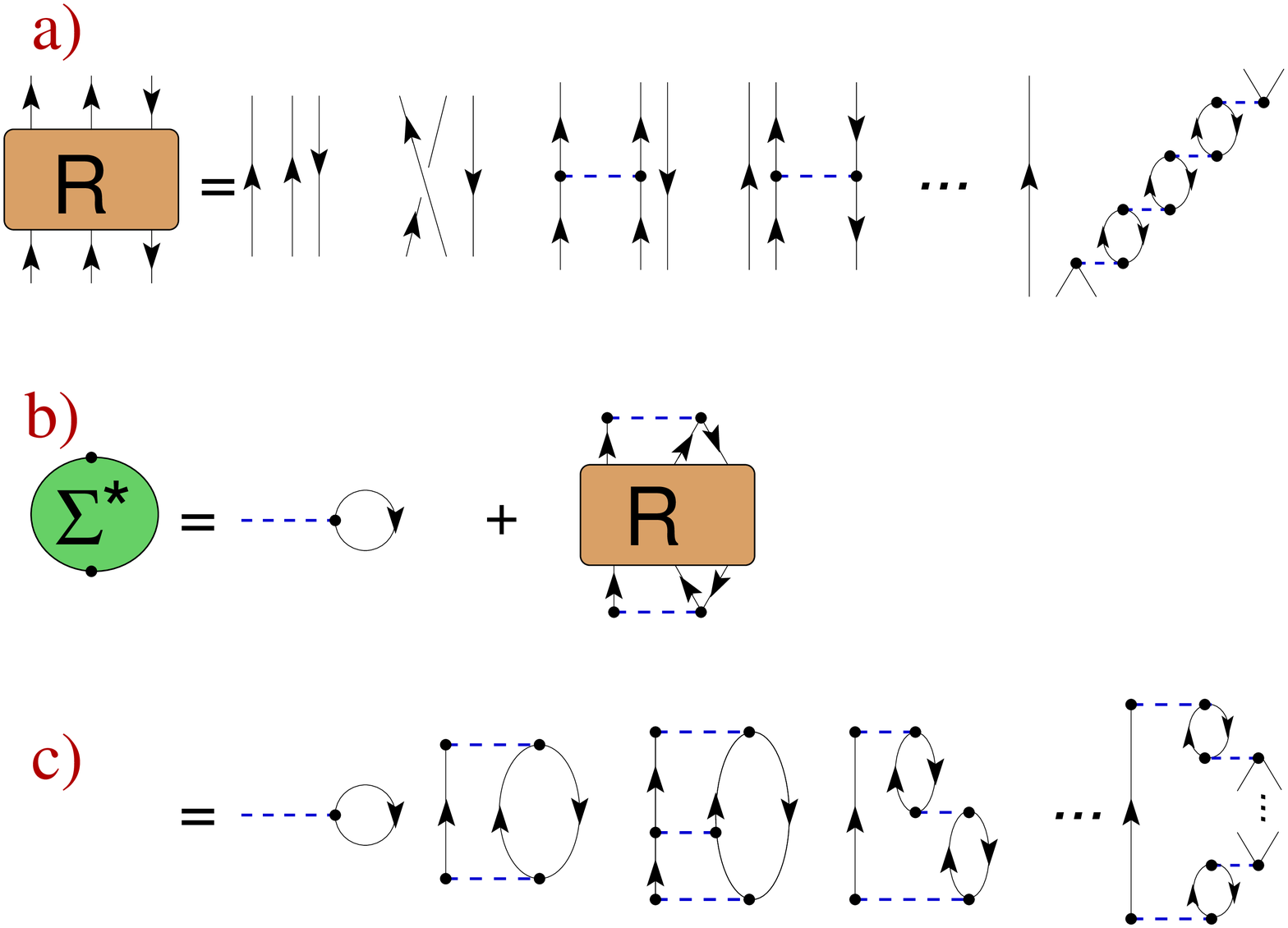}
\caption{(Color online) a)~Diagrammatic expansion of $R(\omega)$ in terms of 
the (antisymmetrized) Coulomb interaction and undressed propagators.
 b)~$R(\omega)$ is related to the self-energy according to 
Eq.~(\ref{eq:Sigma1}). c)~By substituting the diagrams a) in the latter 
equation, one finds the perturbative expansion of the self-energy. }
\label{fig:SigR}
\end{figure}

The sp propagator solves the Dyson equation
\begin{equation}
 g_{\alpha \beta}(\omega) =  g^{0}_{\alpha \beta}(\omega) \; +  \;
 %\sum_{\gamma \delta}
 g^{0}_{\alpha \gamma}(\omega) 
     \Sigma^\star_{\gamma \delta}(\omega)   g_{\delta \beta}(\omega) \; \; ,
\label{eq:Dys}
\end{equation}
which depends on the irreducible self-energy $\Sigma^\star(\omega)$. The 
latter can be written as the sum of two terms
\begin{equation}
 \Sigma^{\star}_{\alpha \beta}(\omega) ~=~  \Sigma^{HF}_{\alpha \beta} 
   ~+~ \frac{1}{4} \, 
  %\sum_{\mu \nu \lambda \\ \gamma \delta \varepsilon}
  V_{\alpha \lambda , \mu \nu}
  ~ R_{\mu \nu \lambda , \gamma \delta \varepsilon}(\omega)
    ~ V_{\gamma \delta , \beta \varepsilon}  \; ,
\label{eq:Sigma1}
\end{equation}
where $\Sigma^{HF}$ represents the Hartree-Fock diagram for the self-energy.
In Eqs.~(\ref{eq:Dys}) and~(\ref{eq:Sigma1}), $g^{0}(\omega)$ is the 
sp propagator for the system of noninteracting electrons, whose Hamiltonian 
contains only the kinetic energy and the electron-nucleus attraction. 
The $V_{\alpha \beta , \gamma \delta}$ represent the antisymmetrized matrix 
elements of the interelectron (Coulomb) repulsion. Note that in this work we 
only consider antisymmetrized elements of the interaction, hence, our result 
for the ring summation always compare to the generalized $GW$ approach. 
Equation~(\ref{eq:Sigma1}) introduces the {\em 2p1h}/{\em 2h1p}-irreducible 
propagator $R(\omega)$, which carries the information concerning the coupling 
of sp states to more complex configurations.
Both $\Sigma^\star(\omega)$ and $R(\omega)$ have a perturbative expansion as a 
power series in the interelectron interaction ${\hat V}$.
% Each term of the series also contains a number of non interacting propagators. A self-consistent Green's function approach involves a regrouping of the series, in which these quantities are expressed in terms of the dressed propagator $g_{\alpha \beta}(\omega)$, rather than the noninteracting ones.
Some of the diagrams appearing in the expansion of $R(\omega)$ are depicted 
in Fig.~\ref{fig:SigR}, together with the corresponding contributions to the 
self-energy. Note that already at zero order in $R(\omega)$ (three free lines 
with no mutual interaction) the  second order self-energy is generated.

%%%%%%%%%%%%%%%%%%%%%%%%%%%%%%%%%%%%%%%%%%%%%%%%%%%%%%%%%%%%%%%%%%%%%%%%%%
%%%%%%%%%%%%%%%%%%%%%%%%%%%%%%%%%%%%%%%%%%%%%%%%%%%%%%%%%%%%%%%%%%%%%%%%%

\begin{figure}
\includegraphics[width=0.65\columnwidth,clip=true]{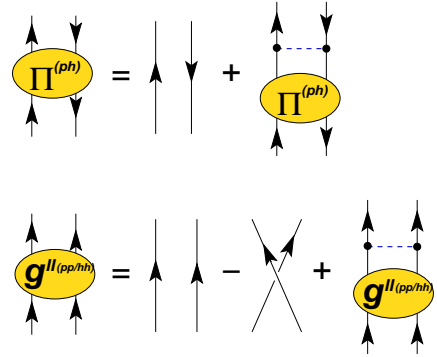}
\caption{(Color online) Diagrammatic equations for the polarization (above) 
and the two-particle (below) propagators in the (G)RPA approach. Dashed lines 
are always antisymmetrized Coulomb matrix elements and the full lines 
represent free (undressed) propagators.}
\label{fig:rpaeq}
\end{figure}

Different approximations to the self-energy can be constructed by summing 
particular classes of diagrams. In this work we are interested in the 
summation of rings and ladders, through the (G)RPA equations.
In order to include such effects in $R(\omega)$, we first consider the 
polarization propagator describing excited states in the $N$-electron system 
\begin{eqnarray}
 \Pi_{\alpha \beta , \gamma \delta}(\omega) &=& 
%% g_{\alpha \beta}(\omega) ~=~ 
 \sum_{n \ne 0}  \frac{  {\mbox{$\langle {\Psi^N_0} \vert $}}
            c^{\dag}_\beta c_\alpha {\mbox{$\vert {\Psi^N_n} \rangle$}} \;
             {\mbox{$\langle {\Psi^N_n} \vert $}}
            c^{\dag}_\gamma c_\delta {\mbox{$\vert {\Psi^N_0} \rangle$}} }
            {\omega - \left( E^N_n - E^N_0 \right) + i \eta } 
\nonumber \\
 &-& \sum_{n \ne 0} \frac{  {\mbox{$\langle {\Psi^N_0} \vert $}}
              c^{\dag}_\gamma c_\delta {\mbox{$\vert {\Psi^N_n} \rangle$}} \;
                 {\mbox{$\langle {\Psi^N_n} \vert $}}
             c^{\dag}_\beta c_\alpha {\mbox{$\vert {\Psi^N_0} \rangle$}} }
            {\omega + \left( E^N_n - E^N_0 \right) - i \eta } \; ,
\label{eq:Pi}
\end{eqnarray}
and the two-particle propagator, that describes the addition/removal of two 
electrons
\begin{eqnarray}
 g^{II}_{\alpha \beta , \gamma \delta}(\omega) &=& 
%% g_{\alpha \beta}(\omega) &=& 
 \sum_n  \frac{  {\mbox{$\langle {\Psi^N_0} \vert $}}
                c_\beta c_\alpha {\mbox{$\vert {\Psi^{N+2}_n} \rangle$}} \;
                 {\mbox{$\langle {\Psi^{N+2}_n} \vert $}}
         c^{\dag}_\gamma c^{\dag}_\delta {\mbox{$\vert {\Psi^N_0} \rangle$}} }
            {\omega - \left( E^{N+2}_n - E^N_0 \right) + i \eta }
\nonumber \\  
&-& \sum_k  \frac{  {\mbox{$\langle {\Psi^N_0} \vert $}}
    c^{\dag}_\gamma c^{\dag}_\delta {\mbox{$\vert {\Psi^{N-2}_k} \rangle$}} \;
                 {\mbox{$\langle {\Psi^{N-2}_k} \vert $}}
                  c_\beta c_\alpha {\mbox{$\vert {\Psi^N_0} \rangle$}} }
            {\omega - \left( E^N_0 - E^{N-2}_k \right) - i \eta } \; .
\label{eq:g2}
\end{eqnarray}
 We note that the expansion of $R(\omega)$ arises from applying the equations 
of motion to the sp propagator~(\ref{eq:g1}), which is associated to the 
ground state~$\vert\Psi^{N}_0\rangle$. Hence, all the Green's functions 
appearing in this expansion will also be ground state based, including 
Eqs.~(\ref{eq:Pi}) and~(\ref{eq:g2}).
 However the latter contain, in their Lehmann representations, all the 
relevant information regarding the excitation of {\em ph} 
and {\em pp/hh} collective modes. The approach of Ref.~\cite{Bar.01} consists
 in computing these quantities by solving the ring-GRPA and the ladder-RPA 
equations~\cite{DicVan}, which are depicted for propagators in 
Fig.~\ref{fig:rpaeq}. In the more general case of a self-consistent 
calculation, a fragmented input propagator can be used and the 
corresponding dressed (G)RPA [D(G)RPA] equations~\cite{DicVan,Geu.94} 
solved [see Eqs.~(\ref{eq:RPA_ph}) and~(\ref{eq:RPA_II})]. Since the 
propagators~(\ref{eq:Pi}) and~(\ref{eq:g2}) reflect two-body correlations, 
they still have to be coupled to an additional sp propagator in order 
to obtain the corresponding
approximation for the {\em 2p1h} and {\em 2h1p} components of $R(\omega)$. 
This is achieved by solving two separate sets of Faddeev equations.

%%%%%%%%%%%%%%%%%%%%%%%%%%%%%%%%%%%%%%%%%%%%%%%%%%%%%%%%%%%%%%%%%%%%%%%%%%
%%%%%%%%%%%%%%%%%%%%%%%%%%%%%%%%%%%%%%%%%%%%%%%%%%%%%%%%%%%%%%%%%%%%%%%%%

 Taking the {\em 2p1h} case as an example, one can split $R^{(2p1h)}(\omega)$ 
in three different components $\bar{R}^{(i)}(\omega)$ ($i=1,2,3$) that differ from 
each other by the last pair of lines that interact in their diagrammatic 
expansion,
\begin{equation}
%\begin{eqnarray}
    \bar{R}^{(2p1h)}_{\alpha \beta \gamma , \mu \nu \lambda}(\omega) =
 \left[ {G^0}^>_{\alpha \beta \gamma , \mu \nu \lambda}(\omega)
        - {G^0}^>_{\beta \alpha \gamma ,  \mu \nu\lambda}(\omega) \right]
  + \sum_{i=1,2,3}     \bar{R}^{(i)}_{\alpha \beta \gamma , \mu \nu \lambda}(\omega)
  \; ,
\label{eq:faddfullR}
%\end{eqnarray}
\end{equation}
where ${G^0}^>(\omega)$ is the {\em 2p1h} propagator for three freely 
propagating lines.
These components are solutions of the following set 
Faddeev equations~\cite{Fad.61}
\begin{eqnarray}
  \lefteqn{
  \bar{R}^{(i)}_{\alpha  \beta  \gamma  ,
           \mu     \nu    \lambda   }(\omega) 
    ~=~ {G^0}^>_{\alpha  \beta  \gamma  ,
                 \mu'    \nu'   \lambda' }(\omega) ~
   \Gamma^{(i)}_{\mu'   \nu'     \lambda'  ,
                 \mu''  \nu''    \lambda'' }(\omega) }
        \hspace{.1in} & &
\nonumber  \\
  & \times & ~
  \left[ \bar{R}^{(j)}_{\mu''   \nu''  \lambda''  ,
                  \mu     \nu    \lambda    }(\omega) ~+~
         \bar{R}^{(k)}_{\mu''   \nu''  \lambda''  ,
                  \mu  \nu    \lambda    }(\omega)
  \right.
\label{eq:FaddTDA}  \\
  & & ~ ~+~ \left.
      {G^0}^>_{\mu''    \nu''   \lambda''   ,
               \mu      \nu     \lambda   }(\omega)
    - {G^0}^>_{\nu''    \mu''   \lambda''   ,
               \mu      \nu     \lambda   }(\omega)
       \right]  \; ,  ~ ~ i=1,2,3 ~ 
%       \right]  \;  ,
\nonumber
\end{eqnarray}
where ($i,j,k$) are cyclic permutations of ($1,2,3$).
The interaction vertices $\Gamma^{(i)}(\omega)$  contain the couplings of 
a {\em ph} or {\em pp/hh} collective excitation and a freely propagating line. 
These are given in the Appendix in terms of the polarization~(\ref{eq:Pi}) and 
two-particle~(\ref{eq:g2}) propagators.
Equations.~(\ref{eq:FaddTDA}) include RPA-like phonons and fully describe the resulting energy dependence of $R(\omega)$. However, they still neglect 
energy-independent contributions--even at low order in the interaction--that 
also correspond to relevant ground-state correlations.
 The latter can be systematically inserted according to
\begin{equation}
  R^{(2p1h)}_{\alpha  \beta  \gamma  ,
              \mu     \nu    \lambda   }(\omega)  ~=~
  U_{\mu     \nu    \lambda , \mu'    \nu'   \lambda' } \;
   \bar{R}^{(2p1h)}_{\mu'   \nu'     \lambda'  ,
             \mu''  \nu''    \lambda''}(\omega)
    \;  U^\dag_{\mu''  \nu''    \lambda'' , \mu     \nu    \lambda} \; ,
\label{eq:URU}
\end{equation}
where $R(\omega)$ is the propagator we employ in Eq.~(\ref{eq:Sigma1}), 
$\bar{R}(\omega)$ is the one obtained by solving Eqs.~(\ref{eq:FaddTDA}),  
${\bf U} \equiv {\bf I} + \Delta{\bf U}$, and $\bf I$ is the identity matrix.
%\begin{equation}
%  U^{(2p1h)}_{\alpha  \beta  \gamma  ,
%           \mu     \nu    \lambda   }  ~=~
%  \delta_{\alpha \mu} \delta_{\beta \nu} \delta_{\gamma \lambda} ~+~
%  U^{(II)}_{\alpha  \beta  \gamma  ,
%           \mu     \nu    \lambda   }  ~+~
%  U^{(\pi)}_{\alpha  \beta  \gamma  ,
%           \mu     \nu    \lambda   }
%\label{eq:Usplit}
%\end{equation}
Following the algebraic diagrammatic construction method~\cite{Wal.81,Sch.83}, 
the energy independent term $\Delta \bf U$ was determined by expanding 
Eq.~(\ref{eq:URU}) in terms of the interaction and imposing that it fulfills 
perturbation theory up to first order (corresponding to third order in the 
self-energy). 
The resulting $\Delta \bf U$, employed in this work, is the same as in 
Ref.~\cite{Wal.81} and is reported in App.~\ref{app_kernel} for completeness. 
It has been shown that the additional diagrams introduced by this correction 
are required to obtain accurate QP properties. 
Equations.~(\ref{eq:FaddTDA}) and~(\ref{eq:URU})
are valid only in the case in which a mean-field 
propagator is used to expand $R(\omega)$. This is the case of the present 
work, which employs Hartree-Fock sp propagators as input. 
The derivation of these 
equations for the general case of a fragmented propagator is given in the 
appendix.
 More details about the actual implementation of the Faddeev formalism 
to {\em 2p1h}/{\em 2h1p} propagation have been presented in
 Ref.~\cite{Bar.01}.  The calculation of the {\em 2h1p} component of
 $R(\omega)$ follows completely analogous steps.
 
%%%%%%%%%%%%%%%%%%%%%%%%%%%%%%%%%%%%%%%%%%%%%%%%%%%%%%%%%%%%%%%%%%%%%%%%%
\begin{figure}[t]
\includegraphics[width=0.7\columnwidth,clip=true]{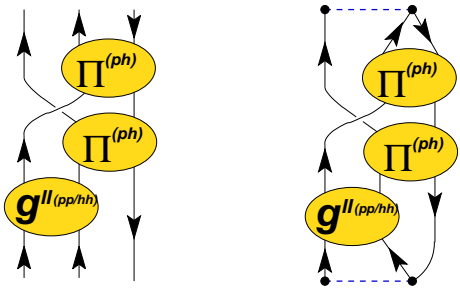}
\caption{(Color online)  Example of one of the diagrams that are summed to all 
orders by means of the Faddeev Eqs.~(\ref{eq:FaddTDA})~(left).
 The corresponding contribution to the self-energy, obtained upon insertion
 into Eq.~(\ref{eq:Sigma1}), is also shown~(right).}
\label{fig:faddex}
\end{figure}
%%%%%%%%%%%%%%%%%%%%%%%%%%%%%%%%%%%%%%%%%%%%%%%%%%%%%%%%%%%%%%%%%%%%%%%%%

It is important to note that the present formalism includes the 
effects of {\em ph} and {\em pp/hh} motion to be included simultaneously, 
while allowing interferences between these modes. 
These excitations
 are evaluated here at the RPA level and are then coupled to each other by 
solving Eqs.~(\ref{eq:FaddTDA}). This generates diagrams as the one displayed 
in Fig.~\ref{fig:faddex}, with the caveat that two phonons are not allowed to 
propagate at the same time. 
Equations.~(\ref{eq:FaddTDA}) also assure that Pauli correlations are properly 
taken into account at the {\em 2p1h}/{\em 2h1p} level. 
In addition, one can in principle
employ dressed sp propagators in these equations to generate a 
self-consistent solution.
 If we neglect the ladder propagator $g^{II}(\omega)$~(\ref{eq:g2}) in this 
expansion, we are left with the ring series alone and the analogous physics 
ingredients as for the generalized $GW$ approach. However, this differs from 
$GGW$ due to the fact that no double counting of the second-order self-energy 
occurs, since the Pauli exchanges between the 
polarization propagator and the third line are properly accounted for (see 
Fig.~\ref{fig:faddex}). Alternatively, one can suppress the polarization 
propagator to investigate the effects of {\em pp/hh} ladders alone.
  
It is instructive to replace in the above equations all RPA phonons with TDA 
ones; this amounts to allowing only forward-propagating diagrams in 
Fig.~\ref{fig:rpaeq}, and is equivalent to 
separate diagonalisations in the spaces of $ph$, $pp$ and $hh$ configurations, 
relative to the HF ground state.
It can be shown that using these TDA phonons to sum all diagrams of the type 
in Fig.~\ref{fig:faddex} reduces to one single diagonalization in the $2p1h$ 
or $2h1p$ spaces. Therefore, Eqs.~(\ref{eq:FaddTDA}) and~(\ref{eq:URU}) with 
TDA phonons lead directly to the ``extended'' $2p1h$ TDA of
 Ref.~\cite{Wal.81}, which was later shown to be equivalent to ADC(3) in the 
general ADC framework~\cite{Sch.83}.
 The Faddeev expansion formalism of Ref.~\cite{Bar.01} creates the possibility 
to go beyond ADC(3) by including RPA phonons. This is more satisfactory in 
the limit of large systems. At the same time, the computational cost remains
 modest since only diagonalizations in the $2p1h/2h1p$ spaces are required. 

Note that complete self-consistency requires the use of fragmented 
(or dressed) propagators in the evaluation of all ingredients leading to the 
self-energy. This is outside the scope of the present paper, but we  
included partial selfconsistency by taking into account the 
modifications to the HF diagram by employing the correlated one-body density 
matrix and iterating to convergence.  
This is relatively simple to achieve, since the $2p1h/2h1p$ propagator is 
only evaluated once with the input HF propagators.
 Below we will give results with and without this 
partial selfconsistency at the HF level.  

%%%%%%%%%%%%%%%%%%%%%%%%%%%%%%%%%%%%%%%%%%%%%%%%%%%%%%%%%%%%%%%%%%%%%%%%%
\begin{table}
%\begin{center}
\begin{ruledtabular}
\begin{tabular}{lcccccccc}
   $l$ &  & 0    & 1   &  2  &  3  &  4  &  5  &  6 \\
  \hline  
 $r_w$ &  &  2.0 & 4.0 & 0.0 & 0.0 & 0.0 & 0.0 & 0.0 \\
 $n_o$ &  &  12  & 21  & 10  & 10  &  5  &  5  &  5 \\
\end{tabular}
\end{ruledtabular}
%\end{center}
 \caption[]{Parameters that define the sp basis: radius of the confining 
wall $r_w$ (in atomic units) and number of orbits $n_o$ used for different 
partial waves $l$. The value of $c_w$ is always set to 5 a.u..}
 \label{tab:basis}
\end{table}
%%%%%%%%%%%%%%%%%%%%%%%%%%%%%%%%%%%%%%%%%%%%%%%%%%%%%%%%%%%%%%%%%%%%%%%%%

\section{Results}
\label{results}

Calculations have been performed using two different model spaces: (1) 
a standard quantumchemical Gaussian basis set, aug-cc-pVTZ for 
Neon~\cite{Dunning}, with Cartesian representation of the $d$ and $f$ 
functions; (2) a numerical basis set based on HF and subsequent 
discretization of the continuum, to be detailed below.    
The aug-cc-pVTZ basis set was used primarily to check our formalism 
with the ADC(3) result in literature (i.e.~\cite{Tro.05}, 
where this basis was 
employed). The HF+continuum basis allows to approach,  
at least for the ionization energies, the results for the 
full sp space (basis set limit).

The HF+continuum is the same discrete model space employed previously 
in Refs.~\cite{Van.01,Ver.06}. It consists of:  
(1) Solving on a radial grid the HF problem for the neutral atom; 
(2) Adding to this fixed nonlocal HF potential a parabolic potential wall 
of the type $U(r)=\theta(r -r_w)c_w(r-r_w)^2$, placed at a distance $r_w$ of 
the nucleus. The latter eigenvalue 
problem has a basis of discrete eigenstates. This basis is truncated 
by specifying some largest angular momentum $l_{\mbox{max}}$ and the number 
of virtual states for each value of $l\leq l_{\mbox{max}}$. (3) Solve the 
HF problem again, without the potential wall, in this truncated discrete 
space. The resulting basis set is used for the subsequent Green's function 
calculations. 

When a sufficiently large number of states is retained after 
truncation, the final results should approach the basis set limit. In 
particular the results should not depend on the choice of the auxiliary 
confining potential. This was verified in Ref.~\cite{Van.01} for the 
second-order, and in Ref.~\cite{Ver.06} for the $G_0W_0$ self-energy; in these 
cases the self-energy is sufficiently simple that extensive convergence 
checks can be made for various choices of the auxiliary potential.     
The parameters of the confining wall and the number of sp 
states kept in the basis set was optimized in Ref.~\cite{Van.01}, 
by requiring that the ionization energy is converged to about 1 mH for the 
second-order self-energy. In Ref.~\cite{Ver.06} the same choice of basis set 
was also seen to bring the ionization energy for the $G_0W_0$ self-energy 
near convergence. For completeness, the details of this basis are reported 
in Table~\ref{tab:basis}.
While the self-energy in the present paper is too complicated to allow similar 
convergence checks, it seems safe to assume that basis set effects will 
affect the calculated ionization energies by at most 5 mH.   

%%%%%%%%%%%%%%%%%%%%%%%%%%%%%%%%%%%%%%%%%%%%%%%%%%%%%%%%%%%%%%%%%%%%%%%%%
\begin{table}[t]
\begin{ruledtabular}
\begin{tabular}{lccccc}
   & F-TDA& F-RPA& F-TDAc&F-RPAc&  Expt. \\
\hline  
2p & -0.799 & -0.791 &-0.803 &-0.797 (0.94) & -0.793 (0.92) \\
2s &  -1.796 &   -1.787  &  -1.802   & -1.793 (0.90)&     -1.782 (0.85) \\
1s &  -32.126  &   -32.087  &  -32.140   &   -32.102 (0.86) & -31.70\\
$E_{\mbox{tot}}$&-128.778&-128.772&-128.836&-128.840&-128.928\\
\end{tabular}
\end{ruledtabular}
%\end{center}
\caption[]{Results with the aug-cc-pVTZ basis. The first three rows   
list the energies of the main sp fragments  
below the Fermi level, as predicted by different 
self-energies.
F-TDA/F-RPA refers to 
the Faddeev summation with TDA/RPA phonons, respectively. 
In all cases the self-energy was corrected at third order through Eq.~(\ref{eq:URU}).
The suffix ``c'' refers to partial selfconsistency, when the static (HF-type) 
self-energy is consistent with the correlated density matrix. Without ``c'' the 
pure HF self-energy was taken. 
In the F-RPAc column the strength of the fragment is indicated 
between brackets. The last row is the total electronic binding energy.  
The experimental values are taken from Refs.~\cite{NIST,Tho.01}. All energies 
are in atomic units.}
\label{tab:2}
\end{table}
%%%%%%%%%%%%%%%%%%%%%%%%%%%%%%%%%%%%%%%%%%%%%%%%%%%%%%%%%%%%%%%%%%%%%%%%%

In Table~\ref{tab:2} we compare, for the aug-cc-pVTZ basis, 
the ionization energies of the main single-hole configurations when 
TDA or RPA phonons are employed in the Faddeev construction (this is labeled 
F-TDA and F-RPA, respectively, in the table). Note that use of 
TDA phonons corresponds to the usual ADC(3) self-energy. We find that 
the replacement of TDA with RPA phonons provides more screening, 
leading to slightly less bound poles which are shifted 
towards the experimental values. This shift increases with binding energy. 
As discussed at the end of Sec.~\ref{intro}, one can 
include consistency of the static part of the self-energy. 
About eight iterations are needed for convergence. 
This is a nonnegligible correction, 
providing about 5 mH more binding (i.e. larger ionization energies)
for the valence/subvalence $2p$ and $2s$, 15 mH for the deeply 
bound $1s$, and 60 mH to the total binding energy.  
Our converged result for the Faddeev-TDA self-energy 
(labeled F-TDAc in Table~\ref{tab:2}) is in good agreement with 
the ADC(3) value for the $2p$ ionization energy (-0.804 H) 
quoted in~\cite{Tro.05}, as it should be. 

The analogous results obtained with the 
larger HF+continuum basis are given in Table~\ref{tab:3},  which allows to    
assess overall stability and basis set effects.  
We find exactly the same trends as for aug-cc-pVTZ. In particular the reduction
of ionization energies from the replacement of TDA with RPA phonons is almost independent of the 
basis set used, while the effect of including partial consistency is roughly halved. 
Overall, the ionization states are always more bound with the larger basis set; 
while the basis set limit could be still more bound than the present  
results with the HF+continuum basis set, it is likely 
(based on the $G_0W_0$ extrapolation in Ref.~\cite{Ver.06}) 
that the difference does not exceed 5 mH.  

%%%%%%%%%%%%%%%%%%%%%%%%%%%%%%%%%%%%%%%%%%%%%%%%%%%%%%%%%%%%%%%%%%%%%%%%%
\begin{table}[t]
\begin{ruledtabular}
\begin{tabular}{lccccc}
   & F-TDA& F-RPA& F-TDAc&F-RPAc&  Expt. \\
\hline  
2p & -0.807 & -0.799 &-0.808 &-0.801 (0.94) & -0.793 (0.92) \\
2s &  -1.802 &   -1.792  &  -1.804   & -1.795 (0.91)&  -1.782 (0.85) \\
1s &  -32.136  &   -32.097  &  -32.142   &   -32.104 (0.81) & -31.70\\
$E_{\mbox{tot}}$&-128.863&-128.857&-128.883&-128.888&-128.928\\
\end{tabular}
\end{ruledtabular}
%\end{center}
\caption[]{Results with the HF+continuum basis set from 
Table~\ref{tab:basis}. 
See also the caption of Table~\ref{tab:2}.} 
\label{tab:3}
\end{table}
%%%%%%%%%%%%%%%%%%%%%%%%%%%%%%%%%%%%%%%%%%%%%%%%%%%%%%%%%%%%%%%%%%%%%%%%%

%%%%%%%%%%%%%%%%%%%%%%%%%%%%%%%%%%%%%%%%%%%%%%%%%%%%%%%%%%%%%%%%%%%%%%%%%
\begin{figure}[t]
\includegraphics[width=0.85\columnwidth,clip=true]{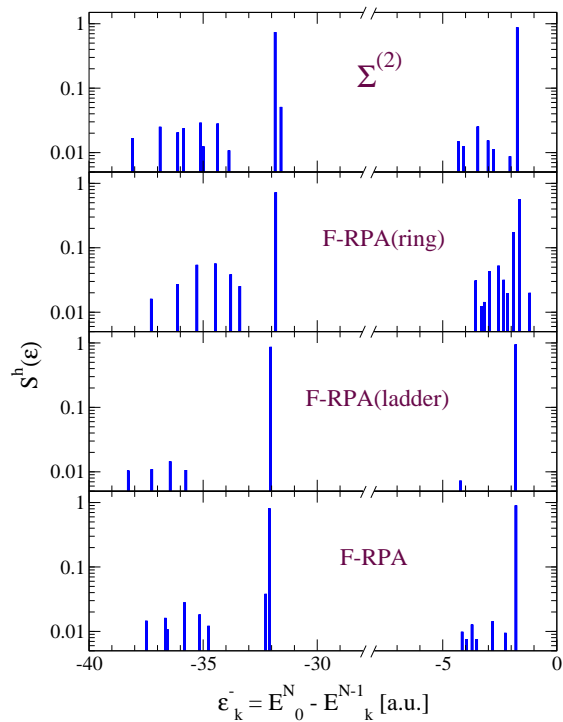}
\caption{(Color online) Spectral function for the $s$ states 
in Ne obtained with various self-energy approximations. From the top down: 
the second-order ($\Sigma^{(2)}$) self-energy, the F-RPA(ring), the 
F-RPA(ladder), and the full F-RPA self-energy. 
The strength is given relative to the Hartree-Fock occupation of each shell. 
Only fragments with strength larger than $Z>0.005$ are shown.
 }
\label{fig:Sh}
\end{figure}
%%%%%%%%%%%%%%%%%%%%%%%%%%%%%%%%%%%%%%%%%%%%%%%%%%%%%%%%%%%%%%%%%%%%%%%%%

%%%%%%%%%%%%%%%%%%%%%%%%%%%%%%%%%%%%%%%%%%%%%%%%%%%%%%%%%%%%%%%%%%%%%%%%%
\begin{table}[t]
\begin{ruledtabular}
\begin{tabular}{lccc}
&1s&2s&2p\\
\hline
HF&-32.77 (1.00) &-1.931 (1.00)&-0.850 (1.00)\\
$\Sigma^{(2)}$&-31.84 (0.74)&-1.736 (0.88)&-0.747 (0.91)\\
$G_0 W_0$&-31.14 (0.85)&-1.774 (0.91)&-0.801 (0.94)\\
F-RPA (ring)&-31.82 (0.73)&-1.636 (0.56)&-0.730 (0.80)\\
F-RPA (ladder)&-32.04 (0.87)&-1.802 (0.95)&-0.781 (0.96)\\
F-RPA&-32.10 (0.81)&-1.792 (0.91)&-0.799 (0.94)\\
Exp.&-31.70&-1.782 (0.85)&-0.793 (0.92)
\end{tabular}
\end{ruledtabular}
%\end{center}
 \caption[]{Energy (in a.u.) and strength (bracketed numbers) of 
the main fragments in the spectral function of Neon, generated by 
different self-energies. 
Results for the HF+continuum basis. Consecutive rows refer to: (1) HF; (2) 
second-order self-energy; (3) $G_0W_0$ results from Ref.~\cite{Ver.06};
(4) F-RPA self-energy with only $ph$ rings retained; (5)  
F-RPA self-energy with only $pp/hh$ ladders retained; (6) Complete 
F-RPA self-energy. 
In all F-RPA results the self-energy was corrected at third order through Eq.~(\ref{eq:URU}).
 The static self-energy was pure HF (no partial self-consistency).   
The experimental values are taken from Refs.~\cite{NIST,Tho.01}. }. 
\label{tab:4}
\end{table}
%%%%%%%%%%%%%%%%%%%%%%%%%%%%%%%%%%%%%%%%%%%%%%%%%%%%%%%%%%%%%%%%%%%%%%%%%

As discussed in Sec.~\ref{intro}, the the F-RPA self-energy contains RPA excitations 
of both $ph$ type (ring diagrams) and $pp/hh$ type (ladder diagrams). 
It is instructive to analyze their separate contributions to the final 
ionization energies, in order to understand how the F-RPA self-energy is 
related to the standard $(G)GW$ self-energy.   
Table~\ref{tab:4} compares the results for the ionization energies, 
obtained with the second-order self-energy, 
to different approximations for including the ring summations. 
 As one can see, the second-order self-energy generates an $l$=1 sp 
energy of -0.747~mH, which is 46~mH above the empirical $2p$ 
ionization energy. 
The $G_0W_0$ self-energy, which includes the ring summation with only direct 
Coulomb matrix elements, improves this result and brings 
it close to experiment. The $2s$ behaves in a similar way. Unfortunately, 
including the exchange terms of the interelectron repulsion in 
the $GG_0W_0$ method  turns out to have the opposite effect 
(the $2p$ ionization energy becomes -0.712 H ~\cite{Ver.06}
\footnote{Note that the $G_0W_0$ and the $GG_0W_0$ results of Ref.~\cite{Ver.06} were obtained by retaining only the 
diagonal part of the electron self-energy,~$\Sigma_{\alpha \alpha}^\star(\omega)$ in the HF+continuum basis.  
This approximation was {\em not} made  in the present work. The error in the ionization 
energies by retaining the diagonal approximation is quite small  (about $\sim$2~mH~\cite{Ver.06} for the Ne atom),  
but larger effects are possible for the total binding energy.})
and the agreement with experiment is lost. 
Obviously, $GG_0W_0$ is too simplistic to account for exchange 
in the $ph$ channel. 
  
With the F-RPA(ring) self-energy one can go one step further and employ 
the Faddeev expansion to also force 
proper Pauli exchange correlations in the {\em 2p1h}/{\em 2h1p} spaces. 
As shown in Table~\ref{tab:4}, 
this enhances the screening due to the exchange interaction terms, leading 
to even less binding for the $2s$ and $2p$.   
The corrections relative to the 
second-order self-energy can be large (100 mH for the $2s$ state) 
and in the direction away from the experimental value. We also note that the 
larger shift, in the 2$s$ orbit, is 
accompanied by an increase of the 
fragmentation~(see Fig.~\ref{fig:Sh} and Tab.~\ref{tab:4}).
Similar observations were also made in Ref.~\cite{Ver.06} for other 
atoms: in general ring summations 
in the direct channel alone bring the quasihole peaks close to the experiment. 
This agreement is 
then spoiled 
as soon as one includes proper exchange terms in the self-energy. 
On the other hand, 
exchange in the {\em ph} channel is required to 
reproduce the correct Rydberg sequence in the excitation spectrum 
of neutral atoms. So 
further corrections must arise from other diagrams, and  obviously the 
summation  of ladder diagrams 
can play a relevant role, since these contribute to the expansion of the 
self-energy at the same level 
as that of the ring diagrams.

The result when only including ladder-type RPA phonons 
in the F-RPA self-energy is also shown in Table~\ref{tab:4}. 
One can see that {\em pp/hh} ladders do actually work in the opposite 
way as the {\em ph} channel ring diagrams, 
and have the same order of magnitude with, {\em e.g.}, a shift of 66 mH 
for the $2s$ relative 
to the second-order result. 
When combined with the ring diagrams in the full F-RPA self-energy, 
the agreement with experiment is restored again.
Note that the final result cannot be obtained by adding the contributions of 
rings and ladders, but depends nontrivially on the interplay between 
these classes of diagrams thereby pointing to significant interference
effects. 

With the F-RPA(ring) self-energy, where only the contributions of the 
$ph$ channel are included, the main peaks listed in Table~\ref{tab:4} 
are not only considerably shifted but also strongly depleted, {\em e.g.} 
a strength of only 0.56 for the main $2s$ peak. 
The complete spectral function for the $l=0$ strength in Fig.~\ref{fig:Sh} 
shows that the depletion of the main fragment is accompanied 
by strong fragmentation over several states. While correlation effects  
are overestimated in F-RPA(ring), they are suppressed in F-RPA(ladder), where 
only the $pp/hh$ ladders are included in the self-energy. In this case one finds a 
spectral distribution closer to the HF one, with a main $2s$ fragment of 
strength 0.95 and less fragmentation than the the second-order self-energy. 
The spectral distribution generated by the complete F-RPA self-energy is
again a combination of the above effects. 
The strength of the deeply bound $1s$ orbital behaves in an analogous way. 
The strength of the main peak is reduced but several satellite levels appear 
due to the mixing with $2h1p$ configurations. In all the calculations 
reported in Fig.~\ref{fig:Sh} we found a summed $l=0$ strength exceeding 0.98 
in the interval [-40 H, -30 H] which can be associated with the $1s$ orbital, 
and this remains true even in the presence of strong correlations using 
the F-RPA(ring) self-energy. Of course, the mixing with $3h2p$ configurations, 
not included in this work, may further contribute to the fragmentation pattern 
in this energy region.  

\section{Conlusions and discussion}
\label{conclusions}

In conclusion, the electronic self-energy for the Ne atom was computed by the F-RPA method which includes  
--simultaneously-- the effects of both ring and ladder diagrams.  
This was accomplished by employing an expansion of the self-energy based on a 
set of Faddeev equations. 
This technique was originally proposed for nuclear structure applications~\cite{Bar.01} and is described in the appendix. 
At the level of the self-energy one sums all diagrams where 
the three propagator lines of the intermediate $2p1h$ or $2h1p$ propagation
are connected by repeated exchange of RPA excitations in both the $ph$ and 
the $pp/hh$ channel. This differs from the ADC(3) formalism in the fact that 
the exchanged excitations are of the RPA type, rather than the TDA type, and 
therefore take ground-state correlations effects into account.  
The coupling to the external points of the self-energy 
uses the same modified vertex as in ADC(3), which must be introduced 
to include consistently all third-order perturbative 
contributions. 

The resulting main ionization energies in the Neon atom are at least of the same 
quality, and even somewhat improved, compared to the ADC(3) result.
Note that, numerically, F-RPA can be implemented as a diagonalization in 
$2p1h/2h1p$ space implying about the same cost as ADC(3).   
The present study also shows that in localized electronic systems 
subtle cancellations occur between the ring and ladder series. 
In particular, only a combination of the ring and ladder series leads to 
sensible results, as the separate ring series tends to correct 
the second-order result in the wrong direction.     
   
Since the limit to extended systems requires an RPA treatment of excitations, 
the F-RPA method holds promise to bridge the gap between an accurate 
description of 
quasiparticles in both finite and extended systems. In particular, the $GW$ 
treatment  
of the electron gas has been shown to yield excellent binding energies, but 
poor quasiparticle properties~\cite{Holm,Godby}. 
Further progress beyond $GW$ theory requires 
a consistent incorporation of  exchange in the $ph$ channel. 
The F-RPA technique  
may be highly relevant in this respect. 
A common framework for calculating accurate QP properties  
in both finite and extended systems, is also important for constraining 
functionals 
in quasiparticle density functional theory (QP-DFT)~\cite{Van.06}.   

Finally, complete self-consistency requires sizable computational efforts for 
bases as large as the HF+continuum basis used here. It would nevertheless 
represent an important extension of the present work, since it is related to 
the fulfillment of conservation laws~\cite{Bay.61,Bay.62}. These issues 
will be addressed in future work.

\acknowledgments This work was supported by the 
U.S. National Science Foundation under grant PHY-0652900.
 
\appendix

\section{Faddeev expansion of the 2p1h/2h1p propagator}
\label{appendix}

Although only the one-energy (or two-time) part of the {\em 2p1h}/{\em 2h1p} propagator enters the definition of the self energy, Eq.~(\ref{eq:Sigma1}), a full resummation of all its diagrammatic contributions would require to treat explicitly the dependence on three separate frequencies, corresponding to the three final lines in the expansion of $R(\omega)$. For example, inserting the RPA ring (ladder) series in $R(\omega)$ implies the propagation of a {\em ph} ({\em pp/hh}) pair of lines both forward and backward in time, while the third line remains unaffected. A way out of this situation is to solve the Bethe-Salpeter-like equations for the polarization and ladder propagators separately and then to couple them to the additional line. If it is assumed that different phonons do not overlap in time, the three lines in between phonon structures will propagate only in one time direction [see figures (\ref{fig:faddex}) and (\ref{fig:int_ex})].
%In other words, one still gives up some time orderings in the full perturbation expansion of $R(\omega)$ but those corresponding to RPA sums of single phonons are completely retained.
%
In this situation the integration over several frequencies can be circumvented following the prescription detailed in the next subsection.
This approach will be discussed in the following for the general case of a fully fragmented propgator, in order to derive a set of Faddeev equations capable of dressing the sp propagator self-consistently.
Since the forward ({\em 2p1h}) and the backward ({\em 2h1p}) parts of $R(\omega)$ decouple in two analogous sets of equations, it is sufficient to focus on the first case alone.

\subsection{Multiple frequencies integrals}
\label{app_time}

%%%%%%%%%%%%%%%%%%%%%%%%%%%%%%%%%%%%%%%%%%%%%%%%%%%%%%%%%%%%%%%%%%%%%%%%%
\begin{figure}[t]
\includegraphics[width=0.4\columnwidth,clip=true]{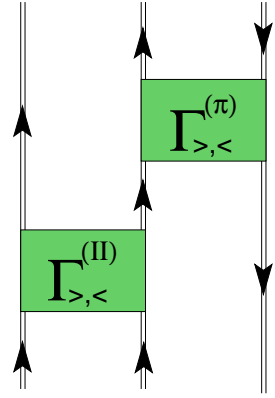}
\caption{(Color online)  Diagrammatic representation of Eq.~(\ref{eq:int_ex}). Double lines represent fully dressed sp Green's funcions which, however, are restricted to propagate only in one time direction [i.e., only one of the two terms on the r.h.s. of Eq.~(\ref{eq:g1}) is retained].
 The Faddeev Eqs.~(\ref{eq:Fadd_1w}) and~(\ref{eq:FaddTDA}) allow for both forward and backward propagation of the phonons $\Gamma^{(\pi)}(\omega)$ and $\Gamma^{(II)}(\omega)$ as long as these do not overlap in time.
 %The diagram on the right represents the last term in Eq.~(\ref{eq:int_ex}), which is associated with the propagatin of {\em 3h2p}. 
For the propagators, time ordereing is asumed with forwad propagation in the upward direction.
}
\label{fig:int_ex}
\end{figure}
%%%%%%%%%%%%%%%%%%%%%%%%%%%%%%%%%%%%%%%%%%%%%%%%%%%%%%%%%%%%%%%%%%%%%%%%%

We start by considering the effective interactions in the {\em ph} and {\em pp}/{\em hh} channels that correspond to Eqs.~(\ref{eq:Pi}) and~(\ref{eq:g2}) stripped of the external legs. In the present work, these are the following two-time objects:
\begin{subequations}
\label{eq:DG}
\begin{eqnarray}
\label{eq:DG_ph}
\Gamma^{(\pi)}_{\alpha \beta, \gamma \delta}(\omega) &=&
  V_{\alpha \delta ,\beta \gamma} ~+~
     V_{\alpha \nu, \beta \mu} 
     \; \Pi^{II}_{\mu \nu, \rho \sigma}(\omega)
     \; V_{\rho \delta, \sigma \gamma}
\\
 &=& V_{\alpha \delta ,\beta \gamma} ~+~
  \sum_{n}  \frac{ \left( \Omega^{n}_{\alpha \beta} \right)^* \;\Omega^{n}_{\gamma \delta} }
                       {\omega - \varepsilon^{\pi}_{n} + i \eta }  ~-~
  \sum_{n'} \frac{ \Omega^{n'}_{\beta \alpha} \; \left( \Omega^{n'}_{\delta \gamma} \right)^*  }
                       {\omega + \varepsilon^{\pi}_{n'} - i \eta } \; ,
\nonumber \\
\label{eq:DG_II}
\Gamma^{(II)}_{\alpha \beta, \gamma \delta}(\omega) &=&
  V_{\alpha \beta, \gamma \delta} ~+~
     V_{\alpha \beta, \mu \nu} 
     \; g^{II}_{\mu \nu, \rho \sigma}(\omega)
     \; V_{\rho \sigma, \gamma \delta}
\\
 &=& V_{\alpha \delta ,\beta \gamma} ~+~
  \sum_n  \frac{ \left( \Delta^{+,n}_{\alpha \beta} \right)^* \;\Delta^{+,n}_{\gamma \delta} }
                       {\omega - \varepsilon^{\Gamma+}_n + i \eta }  ~-~
  \sum_k \frac{ \Delta^{-,k}_{\alpha \beta} \; \left( \Delta^{-,k}_{\gamma \delta} \right)^*  }
                       {\omega - \varepsilon^{\Gamma-}_k - i \eta } \; ,
\nonumber
\end{eqnarray}
\end{subequations}
where the residues and poles for the ring series are
$\Omega^{n}_{\alpha \beta}  =\langle\Psi^N_n     |c^{\dag}_\mu c_\nu        | \Psi^N_0 \rangle V_{\mu \beta, \nu \alpha}$
and $\varepsilon^{\pi}_{n}=E^N_n    -E^N_0$. For the ladders, 
$\Delta^{+,n}_{\alpha \beta}=\langle\Psi^{N+2}_n |c^{\dag}_\mu c^{\dag}_\nu | \Psi^N_0 \rangle V_{\mu \nu, \alpha \beta}$ and
$\Delta^{-,k}_{\alpha \beta}=V_{\alpha \beta, \mu \nu} \langle\Psi^{N-2}_k |c_\mu        c_\nu        | \Psi^N_0 \rangle$,
with poles $\varepsilon^{\Gamma+}_{n}=E^{N+2}_n-E^N_0$ and $\varepsilon^{\Gamma-}_{k}=E^N_0    -E^{N-2}_k$.
Equations.~(\ref{eq:DG}) solve the ring and ladder RPA equations, respectively
\begin{subequations}
\label{eq:RPA}
\begin{eqnarray}
\label{eq:RPA_ph}
 \Gamma^{(\pi)}_{\alpha \beta, \gamma \delta}(\omega) 
 &=&
 V_{\alpha \delta, \beta \gamma}
\\
&+&
 \Gamma^{(\pi)}_{\alpha \beta, \mu \nu}(\omega) 
 \; \int \frac{d \omega_1}{2\pi i}
 g_{\mu \rho}(\omega + \omega_1)  g_{\sigma \nu}(\omega_1)
  \; V_{\rho \delta, \sigma \gamma} \; ,
\nonumber
\\
\label{eq:RPA_II}
\Gamma^{(II)}_{\alpha \beta, \gamma \delta}(\omega)
&=&
  V_{\alpha \beta, \gamma \delta}
\\
&+&
  \Gamma^{(II)}_{\alpha \beta, \mu \nu}(\omega) 
   \; \frac{1}{2} \int \frac{d \omega_1}{-2\pi i}
 g_{\mu \rho}(\omega - \omega_1)  g_{\nu \sigma}(\omega_1)
 \; V_{\rho \sigma, \gamma \delta} 
 \; .
\nonumber
\end{eqnarray}
\end{subequations}

To display how the phonons~(\ref{eq:DG_ph}) and~(\ref{eq:DG_II}) enter the expansion of $R(\omega)$, we perform explicitly the frequency integrals for the diagram of Fig.~\ref{fig:int_ex}. Since it is assumed that the separate propagators lines evolve only in one time direction, only the forwardgoing~($g^>(\omega)$) or the backwardgoing~($g^<(\omega)$) part of Eq.~(\ref{eq:g1}) must be included for particles and holes, respectively. After some algebra, one obtains
\begin{widetext}
\begin{eqnarray}
\lefteqn{
 \Delta R_{\alpha \beta\gamma, \mu \nu \lambda}(\omega)  ~=~
 \int \frac{d \omega_1}{2\pi i} \, \frac{d \omega_2}{2\pi i} \, \frac{d s}{2\pi i} \, \frac{d \Omega}{2\pi i}
}
\label{eq:int_ex} \\
& & \hspace{1cm} 
\; g^>_{\alpha  \alpha_1 }(\omega - \Omega)
\; g^>_{\beta   \beta_1  }(\omega_1)
\; g^<_{\gamma_1 \gamma  }(\omega_1 - \Omega)
\; \Gamma^{(\pi)}_{\beta_1  \gamma_1, \sigma_1 \lambda_1}(\Omega)
\; g^>_{\sigma_1 \sigma_2}(s + \Omega - \omega)
\; \frac{1}{2} \Gamma^{(II)}_{\alpha_1 \sigma_2, \mu_1 \nu_1}(s)
\; g^>_{\mu_1    \mu     }(s - \omega_2)
\; g^>_{\nu_1    \nu     }(\omega_2)
\; g^<_{\lambda \lambda_1}(s - \omega)
\nonumber \\
\nonumber \\
&=& 
\frac{ \left( {\cal X}^{n_1}_{\alpha}   {\cal X}^{n_2}_{\beta}   {\cal Y}^{k_3}_{\gamma} \right)^* \; 
              {\cal X}^{n_1}_{\alpha_1} {\cal X}^{n_2}_{\beta_1} {\cal Y}^{k_3}_{\gamma_1} }
    { \omega - ( \varepsilon^+_{n_1} + \varepsilon^+_{n_2} - \varepsilon^-_{k_3} ) + i \eta }
\;
\left\{
  V_{\beta_1 \lambda_1, \gamma_1 \sigma_1}  ~+~
%\sum_{n_\pi} 
         \frac{\left( \Omega^{n_\pi}_{\beta_1 \gamma_1} \right)^* \Omega^{n_\pi}_{\sigma_1 \lambda_1}  }
              { \omega - ( \varepsilon^+_{n_1} + \varepsilon^{\pi}_{n_\pi}) + i \eta } 
~+~
%\sum_{n'_\pi} 
         \frac{ [ \omega  - \varepsilon^{\pi}_{n'_\pi} - \varepsilon^+_{n_1}   - \varepsilon^+_{n_2}
                          + \varepsilon^-_{k_3}   - \varepsilon^+_{n_4} + \varepsilon^-_{k_7}]
        \;  \Omega^{n'_\pi}_{\gamma_1 \beta_1} \left( \Omega^{n'_\pi}_{\lambda_1 \sigma_1} \right)^*  }
              { [-\varepsilon^{\pi}_{n'_\pi} - \varepsilon^+_{n_2} + \varepsilon^-_{k_3} ]
                [-\varepsilon^{\pi}_{n'_\pi} - \varepsilon^+_{n_4} + \varepsilon^-_{k_7} ] }
 \right\}
\nonumber \\
& & \times ~  
\frac{ \left( {\cal X}^{n_4}_{\sigma_1} \right)^* \; 
              {\cal X}^{n_4}_{\sigma_2} }
    { \omega - ( \varepsilon^+_{n_1} + \varepsilon^+_{n_4} - \varepsilon^-_{k_7} ) + i \eta }
\;
\frac{1}{2}
\left\{
      V_{\alpha_1 \sigma_2, \mu_1 \nu_1}  ~+~
%\sum_{n_{II}} 
         \frac{\left( \Delta^{+,n_{II}}_{\alpha_1 \sigma_2} \right)^* \Delta^{+,n_{II}}_{\mu_1 \nu_1}  }
              { \omega - ( \varepsilon^{\Gamma+}_{n_{II}} - \varepsilon^-_{k_7} ) + i \eta } 
~+~
%\sum_{k_{II}} 
         \frac{ [ \omega  + \varepsilon^{\Gamma-}_{k_{II}} - \varepsilon^+_{n_1}   - \varepsilon^+_{n_4}
                          - \varepsilon^+_{n_5}       - \varepsilon^+_{n_6} + \varepsilon^-_{k_7} ]
        \;  \Delta^{-,k_{II}}_{\alpha_1 \sigma_2} \left( \Delta^{-,k_{II}}_{\mu_1 \nu_1} \right)^*  }
              { [\varepsilon^{\Gamma-}_{k_{II}} - \varepsilon^+_{n_1}  - \varepsilon^+_{n_4} ]
                [\varepsilon^{\Gamma-}_{k_{II}} - \varepsilon^+_{n_5}  - \varepsilon^+_{n_6} ] }
\right\}
\nonumber \\
& &   ~ \times  ~
\frac{ \left( {\cal X}^{n_5}_{\mu_1} {\cal X}^{n_6}_{\nu_1} {\cal Y}^{k_7}_{\lambda_1} \right)^* \; 
              {\cal X}^{n_5}_{\mu}   {\cal X}^{n_6}_{\nu}   {\cal Y}^{k_7}_{\lambda} }
    { \omega - ( \varepsilon^+_{n_5} + \varepsilon^+_{n_6} - \varepsilon^-_{k_7} ) + i \eta }
\nonumber \\
&-& 
\frac{1}{\omega - ( \varepsilon^{\Gamma-}_{k_{II}} - \varepsilon^+_{n_4} - \varepsilon^{\pi}_{n'_\pi} ) - i \eta }
\frac{
 \left( {\cal X}^{n_1}_{\alpha}   {\cal X}^{n_2}_{\beta}   {\cal Y}^{k_3}_{\gamma} \right)^* \; 
              {\cal X}^{n_1}_{\alpha_1} {\cal X}^{n_2}_{\beta_1} {\cal Y}^{k_3}_{\gamma_1}
 ~
 \Omega^{n'_\pi}_{\gamma_1 \beta_1} \left( \Omega^{n'_\pi}_{\lambda_1 \sigma_1} \right)^*
 ~
 \left( {\cal X}^{n_4}_{\sigma_1} \right)^* \;  {\cal X}^{n_4}_{\sigma_2}
 ~
 \Delta^{-,k_{II}}_{\alpha_1 \sigma_2} \left( \Delta^{-,k_{II}}_{\mu_1 \nu_1} \right)^* 
 ~
 \left( {\cal X}^{n_5}_{\mu_1} {\cal X}^{n_6}_{\nu_1} {\cal Y}^{k_7}_{\lambda_1} \right)^* \; 
              {\cal X}^{n_5}_{\mu}   {\cal X}^{n_6}_{\nu}   {\cal Y}^{k_7}_{\lambda} 
}{
    [-\varepsilon^{\pi}_{n'_\pi} - \varepsilon^+_{n_2} + \varepsilon^-_{k_3} ] \; 
    [-\varepsilon^{\pi}_{n'_\pi} - \varepsilon^+_{n_4} + \varepsilon^-_{k_7} ] \;
    [\varepsilon^{\Gamma-}_{k_{II}} - \varepsilon^+_{n_1}  - \varepsilon^+_{n_2} ] \;
    [\varepsilon^{\Gamma-}_{k_{II}} - \varepsilon^+_{n_5}  - \varepsilon^+_{n_6} ]
}  
\nonumber
\end{eqnarray}
\end{widetext}
The last term in this expression contains an energy denominator that involves the simultaneous propagation of two phonons. Thus, it will be discarded in accordance with our assumptions. It must be stressed that similar terms, with overlapping phonons, imply the explicit contribution of at least {\em 3p2h}/{\em 3h2p}. A proper treatment of these would require a non trivial externsion of the present formalism, which is beyond the scope of this paper.

The remaining part in Eq.~(\ref{eq:int_ex}) is the relevant contribution for our purposes. This has the correct energy dependence of a product of denominators that correspond to the intermediate steps of propagation. All of these involve configurations that have at most {\em 2p1h} character. Although, ground state correlations are implicitely included by having already resummed the RPA series. Still, this term does not factorize in a product of separate Green's functions due to the summations over the fragmentation indices $n_i$ and $k_i$ [labeling the eigenstates of the (N$\pm$1)-electron systems]. This is overcome if one defines the matrices ${\bf G^{0>}}(\omega)$, ${\bf \Gamma^{(1,2)}}(\omega)$ and ${\bf \Gamma^{(3)}}(\omega)$, with elements~(no implicit summation used)
\begin{widetext}
\begin{subequations}
\label{eq:G0Gmforward}
\begin{eqnarray}
G^{0>}_{\alpha n_\alpha \beta n_\beta \gamma k_\gamma ; \; \mu n_\mu \nu n_\nu \lambda k_\lambda}(\omega)
 &=& \delta_{n_\alpha , n_\mu} \;  \delta_{n_\beta , n_\nu} \;  \delta_{k_\gamma , k_\lambda} ~
\frac{  \left(   {\cal X}^{n_\alpha}_{\alpha} {\cal X}^{n_\beta}_{\beta} {\cal Y}^{k_\gamma}_{\gamma}  \right)^*
                 {\cal X}^{n_\alpha}_{\mu}    {\cal X}^{n_\beta}_{\nu}   {\cal Y}^{k_\gamma}_{\lambda}\; 
   } { \omega - ( \varepsilon^+_{n_\alpha} + \varepsilon^+_{n_\beta} - \varepsilon^-_{k_\gamma} ) + i \eta }
\; ,
\label{eq:G0f}
\\
\Gamma^{(1)>}_{\alpha n_\alpha \beta n_\beta \gamma k_\gamma ; \; \mu n_\mu \nu n_\nu \lambda k_\lambda}(\omega)
 &=&
\Gamma^{(2)>}_{\beta n_\beta \alpha n_\alpha \gamma k_\gamma ; \; \nu n_\nu \mu n_\mu \lambda k_\lambda}(\omega)
 ~=~
\nonumber \\
 & &
 \hspace{-3cm} = ~
 \frac{\delta_{\alpha ,\mu} \; \delta_{n_\alpha , n_\mu}}{\sum_\sigma \left| {\cal X}^{n_\alpha}_\sigma \right|^2 }
\left\{
  V_{\beta \lambda, \gamma \nu}  ~+~
\sum_{n_\pi} 
         \frac{\left( \Omega^{n_\pi}_{\beta \gamma} \right)^* \Omega^{n_\pi}_{\nu \lambda}  }
              { \omega - ( \varepsilon^+_{n_\alpha} + \varepsilon^{\pi}_{n_\pi}) + i \eta } 
~+~
\sum_{n'_\pi} 
         \frac{ [ \omega  - \varepsilon^{\pi}_{n'_\pi} - \varepsilon^+_{n_\alpha}   - \varepsilon^+_{n_\beta}
                          + \varepsilon^-_{k_\gamma}   - \varepsilon^+_{n_\nu} + \varepsilon^-_{k_\lambda}]
        \;  \Omega^{n'_\pi}_{\gamma \beta} \left( \Omega^{n'_\pi}_{\lambda \nu} \right)^*  }
              { [-\varepsilon^{\pi}_{n'_\pi} - \varepsilon^+_{n_\beta} + \varepsilon^-_{k_\gamma}  ]
                [-\varepsilon^{\pi}_{n'_\pi} - \varepsilon^+_{n_\nu  } + \varepsilon^-_{k_\lambda} ] }
 \right\}
\; ,
\label{eq:Gm12f} \\
\Gamma^{(3)>}_{\alpha n_\alpha \beta n_\beta \gamma k_\gamma ; \; \mu n_\mu \nu n_\nu \lambda k_\lambda}(\omega)
 &=&
\nonumber \\
 & &
 \hspace{-3cm} = ~
\frac{\delta_{\gamma , \lambda} \; \delta_{k_\gamma , k_\lambda}}{ 2 \; \sum_\sigma \left| {\cal Y}^{k_\gamma}_\sigma \right|^2 }
\left\{
      V_{\alpha \beta, \mu \nu}  ~+~
\sum_{n_{II}} 
         \frac{\left( \Delta^{+,n_{II}}_{\alpha \beta} \right)^* \Delta^{+,n_{II}}_{\mu \nu}  }
              { \omega - ( \varepsilon^{\Gamma+}_{n_{II}} - \varepsilon^-_{k_\gamma} ) + i \eta } 
~+~
\sum_{k_{II}} 
         \frac{ [ \omega  + \varepsilon^{\Gamma-}_{k_{II}} - \varepsilon^+_{n_\alpha}   - \varepsilon^+_{n_\beta}
                          - \varepsilon^+_{n_\mu}       - \varepsilon^+_{n_\nu} + \varepsilon^-_{k_\gamma} ]
        \;  \Delta^{-,k_{II}}_{\alpha \beta} \left( \Delta^{-,k_{II}}_{\mu \nu} \right)^*  }
              { [\varepsilon^{\Gamma-}_{k_{II}} - \varepsilon^+_{n_\alpha}  - \varepsilon^+_{n_\beta} ]
                [\varepsilon^{\Gamma-}_{k_{II}} - \varepsilon^+_{n_\mu}     - \varepsilon^+_{n_\nu}   ] }
\right\}
\; . \hspace{1.5cm} 
\label{eq:Gm3f}
\end{eqnarray}
\end{subequations}
\end{widetext}
In these definitions, the row and column indices are ordered to represent at first two quasiparticle lines and then a quasihole. The index `$i$' in $\Gamma^{(i)>}$ refer to the line that propagates independently along with the phonon.
Using Eqs.(\ref{eq:G0Gmforward}), the first term on the r.h.s. of Eq.~(\ref{eq:int_ex}) can be written as
\begin{eqnarray}
\lefteqn{ \Delta R^{(2p1h)}_{\alpha \beta\gamma, \mu \nu \lambda}(\omega)  = } 
& &
\label{eq:factor} \\
& &
\hspace{-0.5cm}
 \sum_{ \tiny \begin{array}{c} n_\alpha \;  n_\beta  \; k_\gamma \\ n_\mu \;  n_\nu \; k_\lambda \end{array}}
%& & \sum_{\frac{n_\alpha \;  n_\beta  \; k_\gamma }{ n_\mu \;  n_\nu \; k_\lambda }}
%& & \sum_{n_\alpha \;  n_\beta  \; k_\gamma \; n_\mu \;  n_\nu \; k_\lambda }
\left[
{\bf G^{0>}}(\omega){\bf \Gamma^{(1)>}}(\omega)
{\bf G^{0>}}(\omega){\bf \Gamma^{(3)>}}(\omega){\bf G^{0>}}(\omega)
\right]_{\alpha n_\alpha \beta n_\beta \gamma k_\gamma ; \; \mu n_\mu \nu n_\nu \lambda k_\lambda}
\nonumber
\end{eqnarray}
Eq.~(\ref{eq:factor}) generalizes to diagrams involving any number of phonon insertions, as long as the terms involving two or more simultaneous phonons are dropped. Based on this relation, we use the following prescription to avoid performing integrals over frequencies. One extends all the Green's functions to objects depending not only on the sp basis' indices ($\alpha,\beta,\gamma$) but also on the indices labeling quasi-particles and holes ($n_i$ and $k_i$). Whether a given argument represents a particle or an hole depends on the type of line being propagated. At this point one can perform calculations working with only two-time quantities. The standard propagator is recovered at the end by summing the ``extended'' one over the quasi-particle/hole indices.

\subsection{Faddeev expansion}
\label{app_faddeqs}

The 2p1h/2h1p propagator that includes the full resummation of both the ladder and ring diagrams at the (G)RPA level is the solution of the following Bethe-Salpeter-like equation,
\begin{eqnarray}
\label{eq:R_BSE}
\lefteqn{
 R_{\alpha \beta\gamma, \mu \nu \lambda}(\omega_1, \omega_2, \omega_3) ~=~
}
 & &
\\
 & & \hspace{-0.8cm}
 = ~ \left[ g_{\alpha \mu}(\omega_1)  g_{\beta \nu}(\omega_2) ~-~ g_{\beta \mu}(\omega_2)  g_{\alpha \nu}(\omega_1) \right] \; g_{\lambda \gamma}(-\omega_3) ~ +
\nonumber \\
& & \hspace{-0.8cm}
\left\{
  g_{\beta \beta_1}(\omega_2)  g_{\gamma_1 \gamma}(-\omega_3)
   V_{\beta_1 \sigma, \gamma_1 \rho} 
  \int \frac{d s}{2\pi i} R_{\alpha \rho \sigma, \mu \nu \lambda}(\omega_1, s, \omega_2 + \omega_3 - s)
\right.  
\nonumber \\
& & \hspace{-0.8cm}
+
\left.
 g_{\alpha \alpha_1}(\omega_1)  g_{\gamma_1 \gamma}(-\omega_3)
   V_{\alpha_1 \sigma, \gamma_1 \rho} 
  \int \frac{d s}{2\pi i} R_{\rho \beta \sigma, \mu \nu \lambda}(s, \omega_2, \omega_1 + \omega_3 - s)
\right.  
\nonumber \\
& & \hspace{-1cm}
+
\left.
 \frac{1}{2}  g_{\alpha \alpha_1}(\omega_1)  g_{\beta \beta_1}(\omega_2)
    V_{\alpha_1 \beta_1, \rho \sigma} 
   \int \frac{d s}{-2\pi i} R_{\rho \sigma\gamma, \mu \nu \lambda}(s, \omega_1 + \omega_2 - s, \omega_3)
\right\}  .
\nonumber
\end{eqnarray}
If this equation is solved, a double integration of $R(\omega_1,\omega_2,\omega_3)$ would yield the two-time propagator $R(\omega)$ contributing to Eq.~(\ref{eq:Sigma1}). However, the numerical solution of Eq.~(\ref{eq:R_BSE}) appears beyond reach of the present day computers and one needs to avoid dealing directly with multiple frequencies integrals. The strategy used is to first solve the RPA equations~(\ref{eq:RPA_ph}) and~(\ref{eq:RPA_II}) separately. Once this is done it is necessary to rearrange the series~(\ref{eq:R_BSE}) in such a way that only the resummed phonons appear. Following the formalism introduced by Faddeev~\cite{Fad.61,Joa.75}, we identify the components $R^{(i)}(\omega)$ with the three terms between curly brakets in Eq.~(\ref{eq:R_BSE}). By employing Eqs.~(\ref{eq:RPA_ph}) and~(\ref{eq:RPA_II}) one is lead to the following set of equations~%
\footnote{Note that the present definitions of the $R^{(i)}$ differ from the ones of Ref.~\cite{Bar.01} which contain the additional term $\frac{1}{2}[G^0-G^{0,ex}]$. The two different forms of the Faddeev equations that result can be easily related into each other and are completely equivalent. The definition used here agrees with the standard literature on the subject~\cite{Fad.61,Joa.75}.},
\begin{eqnarray}
\label{eq:Fadd_3w}
\lefteqn{
 R^{(i)}_{\alpha \beta \gamma, \mu \nu \lambda}(\omega_1, \omega_2, \omega_3) ~=~
 g_{\alpha \alpha_1}(\omega_1)  g_{\beta \beta_1}(\omega_2)  g_{\gamma_1 \gamma}(-\omega_3)
}
 & &
\\
& & \times
 \int \frac{d s_1 \; d s_2 \; d s_3}{2\pi i}
  \Gamma^{(i)}_{\alpha_1 \beta_1 \gamma_1 , \mu_1 \nu_1 \lambda_1}(\omega_1, \omega_2, \omega_3; s_1,s_2, s_3)
\nonumber \\
& & \times
 \left\{
   \left[ g_{\mu_1 \mu}(s_1)  g_{\nu_1 \nu}(s_2)
      ~-~ g_{\nu_1 \mu}(s_2)  g_{\mu_1 \nu}(s_1) \right] \; g_{\lambda \lambda_1}(-s_3)
  \right.
\nonumber \\
& &
 \left.
  + R^{(j)}_{\mu_1 \nu_1 \lambda_1, \mu \nu \lambda}(s_1,s_2, s_3)
  + R^{(k)}_{\mu_1 \nu_1 \lambda_1, \mu \nu \lambda}(s_1,s_2, s_3)
  \right\}
 \; , \; i=1,2,3 \; ,
\nonumber 
\end{eqnarray}
where (i,j,k) are cyclic permutations of (1,2,3) and the interaction vertices $\Gamma^{(i)}(\omega_1,\omega_2,\omega_3)$ are given by
\begin{subequations}
\label{eq:Gamma_3w}
\begin{eqnarray}
 \Gamma^{(1)}_{\alpha \beta \gamma , \mu \nu \lambda}(\omega_1, \omega_2, \omega_3; \omega_4, \omega_5, \omega_6) &=&
\label{eq:Gamma12_3w} \\
 & & \hspace{-5cm} 
 =~ \Gamma^{(2)}_{\beta \alpha \gamma , \nu \mu \lambda}(\omega_2, \omega_1, \omega_3; \omega_5, \omega_4, \omega_6) ~=~
\nonumber \\
 & & \hspace{-5cm} 
 =~ \delta(\omega_1 -\omega_4) \delta(\omega_2 + \omega_3 - \omega_5 - \omega_6)
 g_{\alpha \mu}^{-1}(\omega_1) \Gamma^{(\pi)}_{\beta \gamma, \nu \lambda }(\omega_2 + \omega_3)
 \; ,
\nonumber \\
 \Gamma^{(3)}_{\alpha \beta \gamma , \mu \nu \lambda}(\omega_1, \omega_2, \omega_3; \omega_4, \omega_5, \omega_6) &=&
\label{eq:Gamma3_3w} \\
 & & \hspace{-5cm} 
 = \frac{1}{2} \delta(\omega_3 -\omega_6) \delta(\omega_1 + \omega_2 - \omega_4 - \omega_5)
 g_{\lambda \gamma}^{-1}(-\omega_3) \Gamma^{(II)}_{\alpha \beta, \mu \nu}(\omega_1 + \omega_2)
 \; .
\nonumber 
\end{eqnarray}
\end{subequations}
Finally, we apply the prescription of Sec.~\ref{app_time} and substitute $R(\omega_1,\omega_2,\omega_3)$ with its extended but two-time version $R(\omega)$. This leads to the following set of Faddeev equations which propagate 2p1h forward in time,
\begin{eqnarray}
\label{eq:Fadd_1w}
  \lefteqn{
  \bar{R}^{(i)}_{\alpha n_\alpha \beta n_\beta \gamma  k_\gamma ; \; 
           \mu    n_\mu    \nu   n_\nu   \lambda k_\lambda}(\omega) ~=~ }
     & &
\nonumber  \\
  & & +~ {G^0}^>_{\alpha n_\alpha \beta n_\beta \gamma  k_\gamma ; \; 
                  \alpha' n'_\alpha \beta' n'_\beta \gamma'  k'_\gamma}(\omega) ~
    \Gamma^{(i)}_{\alpha' n'_\alpha \beta' n'_\beta \gamma'  k'_\gamma ; \; 
                  \mu' n_\mu' \nu' n_\nu'  \lambda' k_\lambda'}(\omega) ~
\nonumber  \\
  & &  \times ~
  \left[ {G^0}^>_{\alpha n_\alpha \beta n_\beta \gamma  k_\gamma ; \; 
                  \mu    n_\mu    \nu   n_\nu   \lambda k_\lambda}(\omega)
       - {G^0}^>_{\alpha n_\alpha \beta n_\beta \gamma  k_\gamma ; \; 
                  \mu    n_\mu    \nu   n_\nu   \lambda k_\lambda}(\omega)
       \right.
\nonumber  \\
  & & ~ \times ~
  \left. \bar{R}^{(j)}_{\mu' n_\mu' \nu' n_\nu'  \lambda' k_\lambda' ; \; 
                  \mu    n_\mu    \nu   n_\nu   \lambda k_\lambda}(\omega) ~+~
         \bar{R}^{(k)}_{\mu' n_\mu' \nu' n_\nu'  \lambda' k_\lambda' ; \; 
                  \mu    n_\mu    \nu   n_\nu   \lambda k_\lambda}(\omega)
       \right] \; ,
\nonumber  \\
 & & \hspace{4cm} i=1,2,3 \; .
 \end{eqnarray}
Since the full energy dependence is retained in Eq.~(\ref{eq:Fadd_3w}), the self-energy corresponding to its solution, $R(\omega_1,\omega_2,\omega_3)$, is complete up to third order~[see Eq.~(\ref{eq:Sigma1})]. This is no longer the case after the reduction to a two-time propagator. In particular, the approximation that only forward $2p1h$ propagation is allowed between different phonons implies that all diagrams with different time propagation of their external lines are neglected in Eqs.~(\ref{eq:Fadd_1w}). However, these terms are not energy dependent and can be can be reinserted in a systematic way 
a posteriori as in Eq.~(\ref{eq:URU}). In the general case,
\begin{eqnarray}
\label{eq:URUdressed}
  \lefteqn{
  R_{\alpha \beta \gamma , \mu  \nu \lambda }(\omega) ~=~ }
     & &
\\
  & &  {U}^{(2p1h)}_{\alpha \beta \gamma ; \; 
                  \alpha' n'_\alpha \beta' n'_\beta \gamma'  k'_\gamma} ~
    \bar{R}^{(2p1h)}_{\alpha' n'_\alpha \beta' n'_\beta \gamma'  k'_\gamma ; \; 
                  \mu' n_\mu' \nu' n_\nu'  \lambda' k_\lambda'}(\omega) ~
    {U}^{(2p1h)\;\dag}_{\mu' n_\mu' \nu' n_\nu'  \lambda' k_\lambda' ; \; \mu \nu \lambda}
\nonumber 
\end{eqnarray}
and
\begin{equation}
\label{eq:Udressed}
  {U}^{(2p1h)}_{\alpha \beta \gamma ; \; 
                  \mu    n_\mu    \nu   n_\nu   \lambda k_\lambda} =
    \delta_{\alpha \mu} \delta_{\beta \nu} \delta_{\gamma \lambda} ~+~ 
    \Delta U^{(2p1h)}_{\alpha \beta \gamma ; \; 
                   \mu    n_\mu    \nu   n_\nu   \lambda k_\lambda} \; ,
\end{equation}
where the correction $\Delta U$ can be determined by comparison with perturbation theory.

The vertices~(\ref{eq:G0Gmforward}), that appear in Eqs.~(\ref{eq:Fadd_1w}), and ${U}^{(2p1h)}$ are expressed in terms of the fully fragmented propagator. Therefore, this approach allows to obtain self-consistent solutions of the sp Green's function~\cite{Bar.02}. Whenever, like in this work, only a mean-field propagator is employed as input there exist a one-to-one correspondence between the fragmentation indices and the sp basis. This is expressed by the relations
${\cal X}^n_\alpha=\delta_{n,\alpha}(1-\delta_{\alpha\in F})$ and
${\cal Y}^k_\alpha=\delta_{k,\alpha}\delta_{\alpha\in F}$, where $F$ represents the set of occupied orbits. In this case, it is possible to drop one set of indices so that Eqs.~(\ref{eq:Fadd_1w}) and~(\ref{eq:URUdressed}) simplify into the form~(\ref{eq:FaddTDA}) and~(\ref{eq:URU}).

\subsection{Faddeev vertices}
\label{app_kernel}

In practical applications, it is worth to note that the poles of the free propagator $G^0(\omega)$, Eq.~(\ref{eq:G0f}), do not contribute to the kernel of Eqs.~(\ref{eq:Fadd_1w}). This can be proven by employing the closure relations for the RPA problem, in the form obtained by extracting the free poles in Eqs.~(\ref{eq:RPA}). As an example, for the forward poles of the ladder propagator these are
\begin{eqnarray}
\lefteqn{
  \lim_{\omega \to \varepsilon^+_{n_1} + \varepsilon^+_{n_2}}
   [(\omega - \varepsilon^+_{n_1} - \varepsilon^+_{n_2}) ~ \times ~ \mbox{(Eq.~\ref{eq:RPA_II})]}
 ~ ~\Longrightarrow
 }
 & &
\\
&~&
 \left( {\cal X}^{n_1}_{\alpha}   {\cal X}^{n_2}_{\beta}  \right)^*
              {\cal X}^{n_1}_{\mu} {\cal X}^{n_2}_{\nu}  \; 
      \Gamma^{(II)}_{\mu \nu, \gamma \delta}(\omega = \varepsilon^+_{n_1} + \varepsilon^+_{n_2})
  ~=~ 0 \; , ~ ~ \forall n_1 , n_2 \; ,
\nonumber
\end{eqnarray}
and similarly for other cases.  Making use of these relations one can derive the following working expression of the kernels of the {\em 2p1h} Faddeev equations (no implicit summations used)
\begin{widetext}
\begin{subequations}
\label{eq:vert2p1h}
\begin{eqnarray}
\label{eq:vert2p1h_12}
\left[
  {\bf G^{0>}}(\omega){\bf \Gamma^{(1)>}}(\omega)
\right]_{\alpha n_\alpha \beta n_\beta \gamma k_\gamma ; \; \mu n_\mu \nu n_\nu \lambda k_\lambda} &=&
\left[
  {\bf G^{0>}}(\omega){\bf \Gamma^{(2)>}}(\omega)
\right]_{\beta n_\beta \alpha n_\alpha \gamma k_\gamma ; \; \nu n_\nu \mu n_\mu \lambda k_\lambda} ~=~
\\
& & \hspace{-5cm}
=~ \delta_{n_\alpha, n_\mu} \;
   \frac{\left(  {\cal X}^{n_\alpha}_{\alpha} {\cal X}^{n_\beta}_{\beta} {\cal Y}^{k_\gamma}_{\gamma}  \right)^*
                 {\cal X}^{n_\alpha}_{\mu}  }  %%%%  {\cal X}^{n_\beta}_{\nu}   {\cal Y}^{k_\gamma}_{\lambda}\; }
        { \sum_\sigma \left| {\cal X}^{n_\alpha}_\sigma \right|^2 }
\left\{
\sum_{n_\pi} 
         \frac{ \sum_{\beta_1 \gamma_1} {\cal X}^{n_\beta}_{\beta_1}   {\cal Y}^{k_\gamma}_{\gamma_1}
               \left( \Omega^{n_\pi}_{\beta_1 \gamma_1} \right)^* \Omega^{n_\pi}_{\nu \lambda}  }
              { [\varepsilon^{\pi}_{n_\pi} - \varepsilon^+_{n_\beta} + \varepsilon^-_{k_\gamma}  ]
                [\omega - ( \varepsilon^+_{n_\alpha} + \varepsilon^{\pi}_{n_\pi}) + i \eta         ]} 
~+~
\sum_{n'_\pi} 
         \frac{ \sum_{\beta_1 \gamma_1} {\cal X}^{n_\beta}_{\beta_1}   {\cal Y}^{k_\gamma}_{\gamma_1}
                \Omega^{n'_\pi}_{\gamma_1 \beta_1} \left( \Omega^{n'_\pi}_{\lambda \nu} \right)^*  }
              { [-\varepsilon^{\pi}_{n'_\pi} - \varepsilon^+_{n_\beta} + \varepsilon^-_{k_\gamma}  ]
                [-\varepsilon^{\pi}_{n'_\pi} - \varepsilon^+_{n_\nu  } + \varepsilon^-_{k_\lambda} ]} 
 \right\} \; ,
\nonumber \\
\nonumber \\
\label{eq:vert2p1h_3}
\left[
{\bf G^{0>}}(\omega){\bf \Gamma^{(3)>}}(\omega)
\right]_{\alpha n_\alpha \beta n_\beta \gamma k_\gamma ; \; \mu n_\mu \nu n_\nu \lambda k_\lambda} &=&
\\
& & \hspace{-5cm}
=~ \delta_{k_\gamma, k_\lambda} \;
   \frac{\left(  {\cal X}^{n_\alpha}_{\alpha} {\cal X}^{n_\beta}_{\beta} {\cal Y}^{k_\gamma}_{\gamma}  \right)^*
                 %%%%%{\cal X}^{n_\alpha}_{\mu}    {\cal X}^{n_\beta}_{\nu}  
                 {\cal Y}^{k_\gamma}_{\lambda}\; }
        { 2 \; \sum_\sigma \left| {\cal Y}^{k_\gamma}_\sigma \right|^2 }
\left\{
\sum_{n_{II}} 
         \frac{\sum_{\alpha_1 \beta_1}   {\cal X}^{n_\alpha}_{\alpha_1}    {\cal X}^{n_\beta}_{\beta_1}
               \left( \Delta^{+,n_{II}}_{\alpha_1 \beta_1} \right)^* \Delta^{+,n_{II}}_{\mu \nu}  }
              { [\varepsilon^{\Gamma+}_{n_{II}} - \varepsilon^+_{n_\alpha}  - \varepsilon^+_{n_\beta} ]
                [\omega - ( \varepsilon^{\Gamma+}_{n_{II}} - \varepsilon^-_{k_\gamma} ) + i \eta      ]} 
~+~
\sum_{k_{II}} 
         \frac{\sum_{\alpha_1 \beta_1}   {\cal X}^{n_\alpha}_{\alpha_1}    {\cal X}^{n_\beta}_{\beta_1}
               \Delta^{-,k_{II}}_{\alpha_1 \beta_1} \left( \Delta^{-,k_{II}}_{\mu \nu} \right)^*  }
              { [\varepsilon^{\Gamma-}_{k_{II}} - \varepsilon^+_{n_\alpha}  - \varepsilon^+_{n_\beta} ]
                [\varepsilon^{\Gamma-}_{k_{II}} - \varepsilon^+_{n_\mu}     - \varepsilon^+_{n_\nu}   ] }
\right\} \; .
\nonumber
\end{eqnarray}
\end{subequations}
After substituting Eq.~(\ref{eq:URUdressed}) into (\ref{eq:Sigma1}), one needs the working expression for the matrix product $\bf V \, U^{(2p1h)}$~(where $\bf V$ is the interelectron interaction). The minimum correction that guaranties to reproduce all third order self-energy diagrams is
\begin{eqnarray}
\label{eq:ADC3vt2p1h}
\left[
  {\bf V} \; {\bf U^{(2p1h)}}
\right]_{\alpha ; \; \mu n_\mu \nu n_\nu \lambda k_\lambda} &=&
%  V_{\alpha \lambda, \mu \nu} ~+~
%\left[
%   {\bf V}{\Delta \bf U^{(2p1h)}}
%\right]_{\alpha; \; \mu n_\mu \nu n_\nu \lambda k_\lambda} ~=~
%\\
%& & \hspace{-5cm}
%=~
V_{\alpha \lambda, \mu \nu} ~+~ 
 \frac{ V_{\alpha \lambda, \gamma_1 \delta_1} \; {\cal Y}^{k_\gamma}_{\gamma_1} {\cal Y}^{k_\delta}_{\delta_1}
 \left(   {\cal Y}^{k_\gamma}_{\gamma_2} {\cal Y}^{k_\delta}_{\delta_2} \right)^*  \; V_{\gamma_2 \delta_2, \mu \nu}}
 { 2 \; [\varepsilon^-_{k_\gamma} + \varepsilon^-_{k_\delta} - \varepsilon^+_{n_\mu} - \varepsilon^+_{n_\nu} ]}
%\nonumber
\\
& &~~+~
 \frac{ V_{\alpha \delta_1, \mu \gamma_1} \; {\cal Y}^{k_\gamma}_{\gamma_1} {\cal X}^{n_\delta}_{\delta_1}
 \left(   {\cal Y}^{k_\gamma}_{\gamma_2} {\cal X}^{n_\delta}_{\delta_2} \right)^*  \; V_{\gamma_2 \lambda, \delta_2 \nu}}
 { [\varepsilon^-_{k_\gamma} + \varepsilon^-_{k_\lambda} - \varepsilon^+_{n_\delta} - \varepsilon^+_{n_\nu} ]}
 ~ - ~
 \frac{ V_{\alpha \delta_1, \nu \gamma_1} \; {\cal Y}^{k_\gamma}_{\gamma_1} {\cal X}^{n_\delta}_{\delta_1}
 \left(   {\cal Y}^{k_\gamma}_{\gamma_2} {\cal X}^{n_\delta}_{\delta_2} \right)^*  V_{\gamma_2 \lambda, \delta_2 \mu}}
 { [\varepsilon^-_{k_\gamma} + \varepsilon^-_{k_\lambda} - \varepsilon^+_{n_\delta} - \varepsilon^+_{n_\mu} ]}
 \; .
\nonumber
\end{eqnarray}

The case of {\em 2h1p} is handled in a completely analogous way along the steps of Secs.~(\ref{app_time}) and~(\ref{app_faddeqs}). After extending $R(\omega_1,\omega_2,\omega_3)$ to depend on the fragmentation indices ($k_1$,$k_2$,$n$), the {\em 2h1p} equivalent of Eq.~(\ref{eq:Fadd_1w}) is obtained with the following definitions of the kernels,
\begin{subequations}
\label{eq:vert2h1p}
\begin{eqnarray}
\label{eq:vert2h1p_12}
\left[
  {\bf G^{0>}}(\omega){\bf \Gamma^{(1)>}}(\omega)
\right]_{\alpha k_\alpha \beta k_\beta \gamma n_\gamma ; \; \mu k_\mu \nu k_\nu \lambda n_\lambda} &=&
\left[
  {\bf G^{0>}}(\omega){\bf \Gamma^{(2)>}}(\omega)
\right]_{\beta k_\beta \alpha k_\alpha \gamma n_\gamma ; \; \nu k_\nu \mu k_\mu \lambda n_\lambda} ~=~
\\
& & \hspace{-5cm}
=~ \delta_{k_\alpha, k_\mu} \;
   \frac{        {\cal Y}^{k_\alpha}_{\alpha} {\cal Y}^{k_\beta}_{\beta} {\cal X}^{n_\gamma}_{\gamma} 
         \left(  {\cal Y}^{k_\alpha}_{\mu}  \right)^* }  %%%%  {\cal Y}^{k_\beta}_{\nu}   {\cal X}^{n_\gamma}_{\lambda}\; \right)^* }
        { \sum_\sigma \left| {\cal Y}^{k_\alpha}_\sigma \right|^2 }
\left\{
\sum_{n'_\pi} 
         \frac{ \sum_{\beta_1 \gamma_1} \left( {\cal Y}^{k_\beta}_{\beta_1}   {\cal X}^{n_\gamma}_{\gamma_1} \right)^*
               \Omega^{n'_\pi}_{\gamma_1 \beta_1} \left( \Omega^{n'_\pi}_{\lambda \nu}  \right)^*}
              { [-\varepsilon^{\pi}_{n'_\pi} - \varepsilon^-_{k_\beta} + \varepsilon^+_{n_\gamma}  ]
                [ \omega - ( \varepsilon^-_{k_\alpha} - \varepsilon^{\pi}_{n'_\pi}) - i \eta         ]} 
~+~
\sum_{n_\pi} 
         \frac{ \sum_{\beta_1 \gamma_1} \left(  {\cal Y}^{k_\beta}_{\beta_1}   {\cal X}^{n_\gamma}_{\gamma_1} 
                \Omega^{n_\pi}_{\beta_1 \gamma_1} \right)^* \Omega^{n_\pi}_{\nu \lambda}  }
              { [\varepsilon^{\pi}_{n_\pi} - \varepsilon^-_{k_\beta} + \varepsilon^+_{n_\gamma}  ]
                [\varepsilon^{\pi}_{n_\pi} - \varepsilon^-_{k_\nu  } + \varepsilon^+_{n_\lambda} ]} 
 \right\} \; ,
\nonumber \\
\nonumber \\
\label{eq:vert2h1p_3}
\left[
{\bf G^{0>}}(\omega){\bf \Gamma^{(3)>}}(\omega)
\right]_{\alpha k_\alpha \beta k_\beta \gamma n_\gamma ; \; \mu k_\mu \nu k_\nu \lambda n_\lambda} &=&
\\
& & \hspace{-5cm}
=~ \delta_{n_\gamma, n_\lambda} \;
   \frac{        {\cal Y}^{k_\alpha}_{\alpha} {\cal Y}^{k_\beta}_{\beta} {\cal X}^{n_\gamma}_{\gamma} 
         \left(
            %%%%%{\cal Y}^{k_\alpha}_{\mu}    {\cal Y}^{k_\beta}_{\nu}  
         {\cal X}^{n_\gamma}_{\lambda}\;  \right)^*}
        { 2 \; \sum_\sigma \left| {\cal X}^{n_\gamma}_\sigma \right|^2 }
\left\{
\sum_{k_{II}} 
         \frac{\sum_{\alpha_1 \beta_1} \left(   {\cal Y}^{k_\alpha}_{\alpha_1}    {\cal Y}^{k_\beta}_{\beta_1} \right)^*
               \Delta^{-,n_{II}}_{\alpha_1 \beta_1} \left( \Delta^{-,n_{II}}_{\mu \nu} \right)^*  }
              { [\varepsilon^{\Gamma-}_{k_{II}} - \varepsilon^-_{k_\alpha}  - \varepsilon^-_{k_\beta} ]
                [\omega - ( \varepsilon^{\Gamma-}_{k_{II}} - \varepsilon^+_{n_\gamma} ) - i \eta      ]} 
~+~
\sum_{n_{II}} 
         \frac{\sum_{\alpha_1 \beta_1}  \left(   {\cal Y}^{k_\alpha}_{\alpha_1}    {\cal Y}^{k_\beta}_{\beta_1} 
               \Delta^{+,k_{II}}_{\alpha_1 \beta_1} \right)^* \Delta^{+,k_{II}}_{\mu \nu}   }
              { [\varepsilon^{\Gamma+}_{n_{II}} - \varepsilon^-_{k_\alpha}  - \varepsilon^-_{k_\beta} ]
                [\varepsilon^{\Gamma+}_{n_{II}} - \varepsilon^-_{k_\mu}     - \varepsilon^-_{k_\nu}   ] }
\right\} \; ,
\nonumber
\end{eqnarray}
\end{subequations}
and correction to the external legs,
\begin{eqnarray}
\label{eq:ADC3vt2h1p}
\left[
  {\bf V} \; {\bf U^{(2h1p)}}
\right]_{\alpha ; \; \mu k_\mu \nu k_\nu \lambda n_\lambda} &=&
%  V_{\alpha \lambda, \mu \nu} ~+~
%\left[
%   {\bf V} \; {\Delta \bf U^{(2h1p)}}
%\right]_{\alpha; \; \mu k_\mu \nu k_\nu \lambda n_\lambda} ~=~
%\\
%& & \hspace{-5cm}
%=~
V_{\alpha \lambda, \mu \nu} ~+~ 
 \frac{ V_{\alpha \lambda, \gamma_1 \delta_1} \; {\cal X}^{n_\gamma}_{\gamma_1} {\cal X}^{n_\delta}_{\delta_1}
 \left(   {\cal X}^{n_\gamma}_{\gamma_2} {\cal X}^{n_\delta}_{\delta_2} \right)^*  \; V_{\gamma_2 \delta_2, \mu \nu}}
 { 2 \; [\varepsilon^-_{k_\mu} + \varepsilon^-_{k_\nu} - \varepsilon^+_{n_\gamma} - \varepsilon^+_{n_\delta} ]}
%\nonumber
\\
& &~~+~
 \frac{ V_{\alpha \delta_1, \mu \gamma_1} \; {\cal X}^{n_\gamma}_{\gamma_1} {\cal Y}^{k_\delta}_{\delta_1}
 \left(   {\cal X}^{n_\gamma}_{\gamma_2} {\cal Y}^{k_\delta}_{\delta_2} \right)^*  \; V_{\gamma_2 \lambda, \delta_2 \nu}}
 { [\varepsilon^-_{k_\delta} + \varepsilon^-_{k_\nu} - \varepsilon^+_{n_\gamma} - \varepsilon^+_{n_\lambda} ]}
 ~ - ~
 \frac{ V_{\alpha \delta_1, \nu \gamma_1} \; {\cal X}^{n_\gamma}_{\gamma_1} {\cal Y}^{k_\delta}_{\delta_1}
 \left(   {\cal X}^{n_\gamma}_{\gamma_2} {\cal Y}^{k_\delta}_{\delta_2} \right)^*  V_{\gamma_2 \lambda, \delta_2 \mu}}
 { [\varepsilon^-_{k_\delta} + \varepsilon^-_{k_\mu} - \varepsilon^+_{n_\gamma} - \varepsilon^+_{n_\lambda} ]}
 \; .
\nonumber
\end{eqnarray}
\end{widetext}

It should be pointed out that while the prescription of Sec.~\ref{app_time} allows sp lines to propagate only in one time direction, it allows for backward propagation of the phonons. These contributions translate directly into the energy independent terms of Eqs.~(\ref{eq:vert2p1h}) and~(\ref{eq:vert2h1p}) and are a direct consequence of the inversion pattern typical of RPA theory. These terms have normally a weaker impact than the direct ones on the solutions of Eqs.~(\ref{eq:Fadd_1w}). However, it is show in Ref.~\cite{Bar.01} that they are crucial to guarantee the exact separation of the spurious solutions---always introduced by the Faddeev formalism~\cite{Adh.79,Eva.81}---if  RPA phonons are used. For the same reasons, the last terms in curly brackets of Eqs.~(\ref{eq:vert2p1h}) and~(\ref{eq:vert2h1p}) should be dropped whenever Tamm-Dancoff (TDA) phonons are propagated.

The approach followed in this work for solving Eqs.~(\ref{eq:Fadd_1w}) is to transform them into a matrix representations~\cite{Bar.01}. Once this is done, one is left with an eigenvalue problem that depends only on the {\em 2p1h}~({\em 2h1p}) configurations ($n,n',k$)~[($k,k',n$)]. The spurious states are known exactly~\cite{Bar.01} and can be projected out analytically to reduce the computational load. In any case, they would give vanishing contributions to Eq.~(\ref{eq:Sigma1}).

\end{document}